\documentclass[%
 reprint,
 amsmath,amssymb,
 aps,
 prl,
 floatfix
]{revtex4-2}

\usepackage{graphicx}% Include figure files
\usepackage{dcolumn}% Align table columns on decimal point
\usepackage{bm}% bold math
\usepackage[dvipsnames]{xcolor}
\usepackage{soul}
\usepackage{lipsum}
\usepackage{subfigure}
\usepackage{lineno}
\newcommand{\comment}[1]{} 

\newcommand{\eqnS}[0]{Eqs.~\eqref{eqn:FKPP} and \eqref{eqn:KPZ} }

\usepackage{color}
\definecolor{shiningblue}{rgb}{0.3,0.68,0.89}

\newcommand{\DS}[1]{\textcolor{black}{#1}}

\begin{document}
%\preprint{APS/123-QED}

\title{New sector morphologies emerge from anisotropic colony growth}

\author{Daniel W. Swartz}
\author{Hyunseok Lee}
\author{Mehran Kardar}
\affiliation{Department of Physics, Massachusetts Institute of Technology, Cambridge, Massachusetts 02139, USA}
\author{Kirill S. Korolev}
\affiliation{Department of Physics, Graduate Program in Bioinformatics and Biological Design Center, Boston University, Boston, Massachusetts 02215, USA}
%\date{\today}

\begin{abstract}
    Competition during range expansions is of great interest from both practical and theoretical view points. Experimentally, range expansions are often studied in homogeneous Petri dishes, which lack spatial anisotropy that might be present in realistic populations. Here, we analyze a model of anisotropic growth, based on coupled Kardar-Parisi-Zhang and Fisher-Kolmogorov-Petrovsky-Piskunov equations that describe surface growth and lateral competition. \DS{The anisotropy is encoded in how strongly genetic boundaries between strains are moved as a result of the expansion front morphology}. We completely characterize spatial patterns and invasion velocities in this generalized model. In particular, we find that strong anisotropy results in a distinct morphology of spatial invasion with a kink in the displaced strain ahead of the boundary between the strains. This morphology of the out-competed strain is similar to a shock wave and serves as a signature of anisotropic growth. \DS{We confirm these predictions with a commonly employed reaction-diffusion model of anisotropic growth.} 
\end{abstract}
%\begin{linenumbers}
\maketitle
\section{Introduction}
Surface growth appears in diverse contexts such as molecular beam epitaxy, flame propagation, imbibing, and many others~\cite{joyce1985molecular, hirschfelder1949theory, dhillon2015critical}. Growth patterns are also relevant to ecology and evolution as they describe population dynamics in epidemics, ecological succession, and the spreading of a biofilm or a cancerous tumor~\cite{murray2002mathematical}. These biological examples may harbor an extra layer of complexity when distinct subpopulations compete with each other during a spatial expansion. Therefore, there is substantial interest in understanding and potentially exploiting the intricate interplay of population growth and competition in space~\cite{hallatschek2007genetic, korolev2015evolution, plummer2019fixation, lee2022slow, gralka2016allele}.

\DS{Historically, anisotropic expansions were first studied in the context of crystal growth.  
For slow, quasi-equilibrium growth, the shape of a growing solid must minimize its surface energy.  
A detailed analysis of this process by Gibbs, Curie, and Wulff \cite{li2016gibbs, markov2016crystal} led to the so-called Wulff construction, which determines the shape of a crystal based on the dependence of surface tension on surface orientation relative to the crystal lattice.  
Thus, the anisotropic shapes of crystals were ultimately explained by their broken rotational symmetry. } 

\DS{A slightly different mechanism explains anisotropy in rapidly growing structures such as snowflakes and crystals formed under rapid growth \cite{langer1989dendrites, ben1993snowflake, jindal2009theoretical}.  
Instead of surface tension, orientation-dependent growth velocities determine the final shape of the growing object and its anisotropy~\cite{adam2005flowers}.  
For crystals and snowflakes, the anisotropy of growth velocities originates from the broken rotational symmetry of the solid phase.  
For biological populations, growth velocities may also reflect the broken rotational symmetry of the environment.  
Indeed, preferred expansion directions can arise due to external factors such as collagen fibers, transport channels, chemical gradients, or even magnetic fields \cite{holmes2015transporting, ahmed2021engineering, stylianopoulos2010diffusion, hong1971magnetic}.  }

\DS{The effects of preferred directions on non-equilibrium growth have been studied in several models, including the Ising model of magnetism and the Eden model of population expansions~\cite{wolf1987wulff, devillard1992kinetic, hirsch1986anisotropy}.  
The latter is a classic model for colony and tumor growth, based on stochastic update rules on a lattice where each site can be either empty or occupied by a cell.  
At each simulation step, an individual is selected to reproduce—typically at random—into one of its unoccupied neighboring sites.  
In the standard isotropic Eden model, colonies grow into approximately circular shapes over time.  
However, modified Eden models incorporating a preferred growth direction generate anisotropic, faceted colonies~\cite{wolf1987wulff, devillard1992kinetic, hirsch1986anisotropy}.  
The extent of each facet is inversely related to the speed at which it expands into empty territory.  }

\DS{For nonliving matter, the primary consequence of anisotropic growth is a non-spherical or non-circular shape.  
In biological populations, however, the composition of the population is often of greater interest—for example, the spatial distribution of various genotypes or subclones.  
These genotypes may differ in properties such as growth rate, motility, and drug resistance, or they may be nearly identical except for a distinct genetic marker that does not affect fitness.  
Such neutral markers are commonly used to visualize internal population dynamics.  
In microbial colonies, for instance, subpopulations can be directly observed by inserting genes that encode fluorescent proteins of different colors~\cite{hallatschek2007genetic}.  }

\DS{As colonies grow, genetic drift at the expansion front rapidly segregates different genotypes into genetically homogeneous sectors \cite{hallatschek2007genetic}.
The motion of the sector boundaries between strains is influenced by the many aspects of each strains phenotype \cite{korolev_genetic_2010, hallatschek_life_2010, lee2022slow} and the direction of sector boundary motion can be used as an assay of fitness during range expansion.
For isotropic growth, the sector boundary between two neutral strains is, on average, perpendicular to the population front~(Fig.~\ref{fig:cartoon-anisotropic}).  
Anisotropic growth, however, breaks this constraint, allowing the boundary to form at an arbitrary angle~$\epsilon$ relative to the normal of the expanding front.  
Moreover, this additional sector tilt can vary with the orientation of the expansion front, as demonstrated in Fig.~\ref{fig:cartoon-anisotropic}.  
}

\DS{The aim of this paper is to characterize the morphology of sector boundaries in the presence of both anisotropy and phenotypic differences between two strains.  
To circumvent the difficulties associated with detailed mechanistic models of anisotropy and microbial competition, we adopt a simplified, analytically tractable framework that considers only the dynamics at the colony's edge.  
This approximation, known as the thin-edge or moving-boundary approximation in physics~\cite{van1998three}, is well-suited for microbial colonies as the actively growing layer is only a few hundred microns thick~\cite{giometto2018physical}, while the bulk of the colony exhibits little growth or competition between cells~\cite{hallatschek2007genetic}.  } 

\DS{For isotropic growth, this framework was developed in Refs.~\cite{swartz2023interplay, horowitz2019bacterial, george_chirality_2018} and consists of two coupled partial differential equations.  
The shape of the colony is described by the position or height of the colony edge, which follows a modified Kardar-Parisi-Zhang~(KPZ) equation~\cite{kardar1986dynamic}, a classic model of surface growth.  
The relative abundance of the competing strains is governed by a modified Fisher-Kolmogorov-Petrovsky-Piskunov~(FKPP) equation~\cite{fisher1937wave, kolmogorov1937moscow}, a well-established model for traveling fronts in chemistry and biology.  
The coupling between these two equations captures how colony shape influences competition and how competition gives rise to characteristic colony morphologies~\cite{swartz2023interplay, horowitz2019bacterial, george_chirality_2018}.  }

\DS{Using this framework, we previously determined how the invasion rate of a mutant depends on its phenomenological parameters, such as expansion velocity and competitive strength~\cite{swartz2023interplay}.  
We also predicted all possible colony shapes that can arise from a mutation at the edge of a growing colony.  
Typically, a successful mutant produces a sector that bulges outward due to its higher growth rate~\cite{korolev2012selective}.  
However, our model also predicted alternative morphologies, such as an inverted bulge—i.e., a dent—which has been observed experimentally in recent studies~\cite{lee2022slow}.  }

\DS{Here, we extend the moving boundary approximation of competition to spatially anisotropic growth and generalize the results of Ref.~\cite{swartz2023interplay}. We find that anisotropy does not result in any new terms in the coupled KPZ and FKPP equations, but relaxes a constraint between two parameters, which have to be equal to each other for the growth to be isotropic. The ratio of these parameters~($\beta$ and $\lambda$ in \eqnS) can then be regarded as a measure of anisotropy. We find that weak anisotropy leads only to quantitative changes in the dynamics, but that once the parameter ratio exceeds a certain threshold, a new sector morphology appears. This morphology is characterized by a ``shock'' wave spreading ahead of the sector established by the fitter strain. We present a complete analytical picture of anisotropic growth and confirm our results with simulations of both the simplified model and a mechanistic reaction-diffusion model of cell growth in two spatial dimensions.}

\section{Model}
\subsection{Thin edge approximation}
In many contexts, two-dimensional colony growth on a substrate is effectively confined to a thin region at the population edge due to nutrient limitations~\cite{giometto_physical_2018, kayser2019collective}, pressure buildup~\cite{giometto_physical_2018, hallatschek_life_2010, beroz2018verticalization}, or other factors.  
As a result, the state of the population can be described by two effectively one-dimensional variables: the position of the front,~$h$, and the relative fraction of the genotype of interest,~$f$.  
Both variables depend on time~$t$ and position along the population edge~$x$.  
For simplicity, we consider nearly planar fronts without overhangs, so that~$x$ uniquely specifies a point on the growing front.  
Since there are no dynamics behind the growing edge, the full two-dimensional population structure of the colony can be reconstructed from~$h(x,t)$ and~$f(x,t)$.   

To construct a dynamical model that depends only on~$f$ and~$h$, we must further assume that growth and competition are determined solely by the local state of the population.  
For example, we restrict our analysis to growth conditions with abundant nutrients; otherwise, growth would be influenced by the spatial distribution of nutrients, which is not accounted for in this modeling framework.  

The FKPP equation~\cite{fisher1937wave, kolmogorov1937moscow, roques2012allee, fife1977approach, van2003front} provides a suitable starting point for describing the invasion of a mutant with relative abundance~$f(x,t)$, initially confined to a small spatial domain:  
\begin{equation}
   \frac{\partial f}{\partial t} = s_0 f (1-f)  + D_f \frac{\partial^2 f}{\partial x^2} + \beta \frac{\partial h}{\partial x}\frac{\partial f}{\partial x}\,.
    \label{eqn:FKPP}
\end{equation}
\DS{On the right-hand side, the first term represents the selective advantage of the mutant strain, while the second describes the diffusion of mutants along the one-dimensional expansion front.  
The third term, first introduced in Refs.~\cite{drossel2000phase, horowitz2019bacterial, george_chirality_2018}, captures the effects of colony morphology (see Fig.~\ref{fig:cartoon-anisotropic}) and is further explained in the next section.}   

\DS{Without coupling~($\beta = 0$) or for a perfectly flat front~($\partial_x h = 0$), Eq.~(\ref{eqn:FKPP}) reduces to the standard model for the spread of an advantageous mutation in a one-dimensional population, such as inside a channel.  
Specifically, the FKPP equation describes how an advantageous mutant~($s > 0$) first establishes locally—i.e.,~$f$ saturates to~$f = 1$ in a small spatial domain—before spreading outward with velocity~$u_0 = 2\sqrt{s D_f}$.}  

The dynamics of~$h$ are governed by the KPZ equation~\cite{kardar1986dynamic}:  
\begin{equation}
    \frac{\partial h}{\partial t} = v + \frac{\lambda}{2} \left(\frac{\partial h}{\partial x} \right)^2 + D_h\frac{\partial^2 h}{\partial x^2}\,.
    \label{eqn:KPZ}
\end{equation}
Here, $v$ is the expansion speed of a flat front, which may depend on~$f$, while the term with $D_h$ accounts for the relaxation of front undulations due to the dependence of expansion velocity on local curvature~\cite{kayser2019collective, giometto2018physical}.  
We use the simplest $f$ dependent velocity, $v(f) = v_0 + \alpha f$ where $\alpha$ is understood as the speed difference between the two strains.
The nonlinear term with parameter~$\lambda$ has geometric origins, as described below.  

\subsection{Geometry of growth and competition}
\begin{figure}
    \centering
    \includegraphics[width=0.45\textwidth]{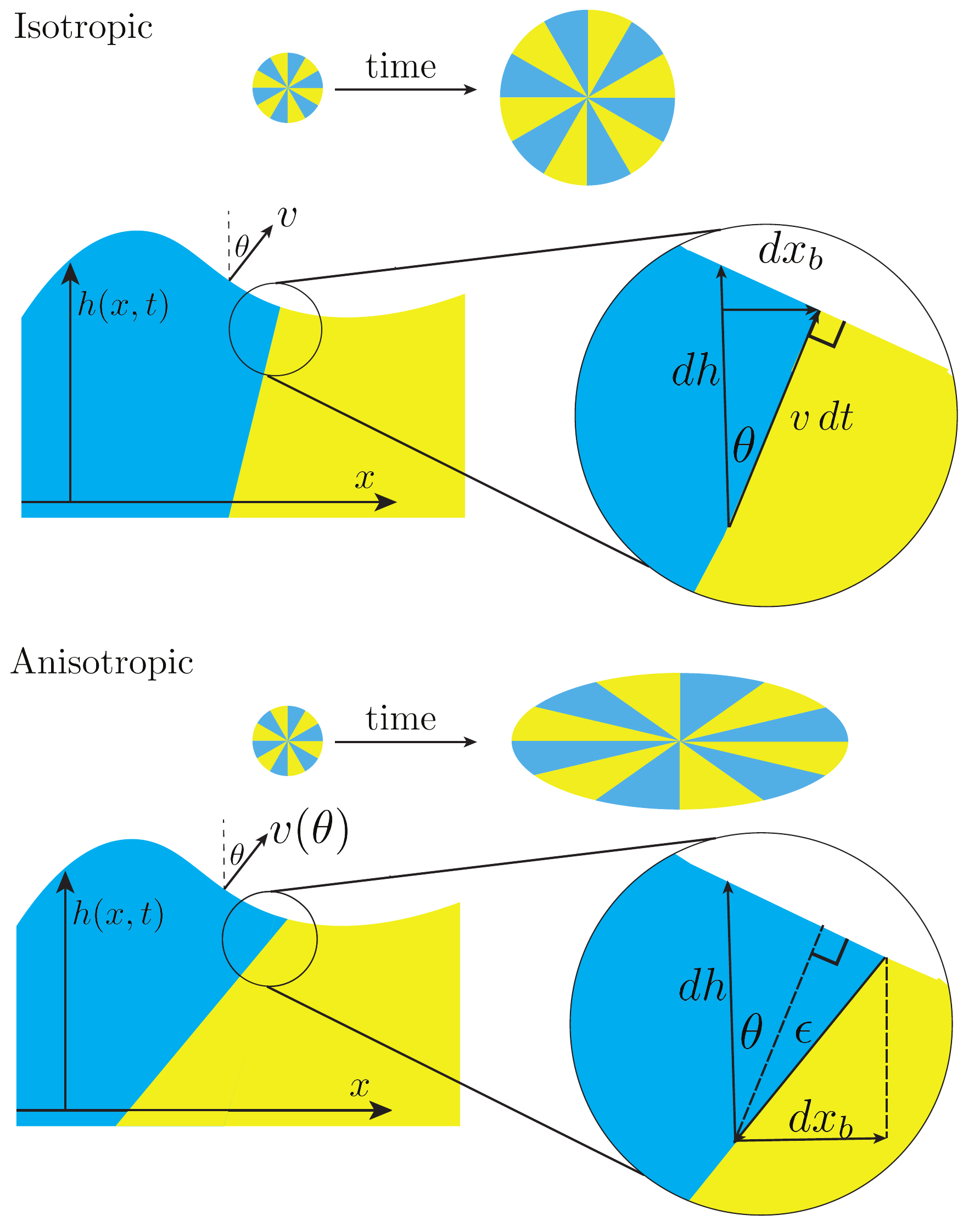}
    \caption{\textbf{Colony growth can be described by a height function~$h(x,t)$ and population composition~$f(x,t)$.} (Color Online)  
    The top panel shows two neutral strains (blue and yellow) with identical expansion speeds growing under isotropic conditions.  
    The resulting colony shape is circular, and the sector boundaries expand perpendicular to the expansion front.  
    The second panel illustrates neutral expansion with a speed~$v(\theta)$ that depends on the growth direction, specified by the angle~$\theta$.  
    As an example, an elliptical colony shape is shown.  
    In general, the boundary between the strains is not orthogonal to the growth front but instead forms an angle~$\epsilon(\theta)$ with the front normal.  
    The vertical growth~$dh$ and horizontal drift~$dx_b$ over a time interval~$dt$ depend on the slope~$\theta$, leading to the nonlinearities in Eqs.~\eqref{eqn:FKPP} and \eqref{eqn:KPZ}.  
    }
    \label{fig:cartoon-anisotropic}
\end{figure}

\DS{We now discuss the geometric origin of both the KPZ nonlinearity~(the $\lambda$-term) and the FKPP coupling~(the $\beta$-term) by considering the dynamics of a neutral sector~($s = 0$); see Fig.~\ref{fig:cartoon-anisotropic}.
It is instructive to first analyze the case of isotropic growth, where the colony front moves at a velocity~$v$ that is independent of the heading angle~$\theta$, i.e., the angle between the vertical and the normal to the growth front.  
During a time interval~$dt$, the height increment~$dh$ is given by~$v\,dt / \cos\theta$, while the demarcation point between the strains drifts by~$dx_b = v\,dt\,\sin\theta$.  
In the limit of small slopes~($\theta \ll 1$), this leads to }
\begin{equation}
    \frac{d h}{d t} = v\left(1 + \frac{\theta^2}{2}\right)\,,\quad{\rm and}\quad \frac{d x_b}{d t} = v\theta \,.
    \label{eqn:dbars}
\end{equation}

\DS{The terms in question then follow directly from the equations above by expressing~$\theta$ as~$- \partial h / \partial x$.  
Specifically, the first equation in \eqref{eqn:dbars} yields the nonlinear term in the KPZ equation with~$\lambda = v$.  }
\DS{The second equation implies that a sector boundary advects, or drifts, with a speed of~$v \partial_x h$. This effect is captured by the advection term added to the standard FKPP equation with~$\beta = v$:}  
\begin{equation}
\left( \frac{\partial f}{\partial t}\right)_{\mathrm{drift}} = -v\theta\frac{\partial f}{\partial x} = v\frac{\partial h}{\partial x}\frac{\partial f}{\partial x}\,.
    \label{eqn:drift}
\end{equation}

\DS{To complete the description of the model studied in Ref.~\cite{swartz2023interplay}, we need to specify how~$v$ in Eq.~(\ref{eqn:KPZ}) depends on~$f$.  
A linear dependence,~$v(f) \approx v_0 + \alpha f$, is the simplest functional form that allows the two types to have different expansion velocities:~$v_0$ and~$v_0 + \alpha$.  
This linear interpolation should be a reasonable assumption in the absence of antagonistic or mutualistic interactions between the types, which would instead be captured by a quadratic~$v(f)$ with a minimum or maximum at intermediate values of~$f$.}  

\DS{Although the~$f$-dependence of~$v$ could be propagated to~$\beta = v$ and~$\lambda = v$, we focus on a simpler model of isotropic growth with~$\lambda = \beta = v_0$, as it remains consistent with the small-angle approximation used in Eqs.~(\ref{eqn:dbars}) to derive the coupled KPZ and FKPP equations.  
Indeed, variations in~$h$ are driven by differences in expansion velocity,~$\alpha$, since a flat colony front would remain flat for~$\alpha = 0$.  
It is straightforward to show that~$\partial h / \partial x$ should be of order~$\alpha$~\cite{swartz2023interplay}, implying that both the nonlinear term in the KPZ equation and the coupling term in the FKPP equation are of order~$\alpha$.  
The~$f$-dependence of~$\lambda$ and~$\beta$ would introduce next-order corrections in~$\alpha$, and it is prudent to omit them from the model, given that similar terms were neglected in Eqs.~(\ref{eqn:dbars}).}  

\subsection{Spatial anisotropy}  

\DS{To account for anisotropy, we repeat the steps used to construct the model above, now considering an expansion velocity that depends on the heading angle~$\theta$.}  

\DS{In typical biaxial anisotropy, an initially circular seed grows into a lozenge shape~\cite{storck2014variable}, with the largest facet advancing along the direction of {\it slowest} growth.  
This behavior is evident in Fig.~\ref{fig:cartoon-anisotropic}, where a small circular colony expands more rapidly along the horizontal direction than the vertical direction, causing most of the colony perimeter to be vertically oriented.}  

\DS{For simplicity, we assume that both species exhibit the same angular dependence of~$v$.  
It is then convenient to align the vertical axis with the direction that minimizes~$v(\theta)$, so that~$\theta = 0$ corresponds to a flat front.  
Expanding~$v(\theta)$ around this slowest direction gives  
\[
v(\theta) \approx v(0) + \frac{1}{2} v''(0) \theta^2.
\]  
To leading order in~$\theta$, the correction~$v''(0) \theta^2 / 2$ contributes only to~$dh/dt$ in Eqs.~(\ref{eqn:dbars}) and, consequently, to the nonlinear term in the KPZ equation.  
Specifically, the value of~$\lambda$ shifts from~$v_0$ to~$v_0 + v''(0)$.}  

\DS{Anisotropy also modifies the parameter~$\beta$ because it affects the motion of sector boundaries, as shown in Fig.~\ref{fig:cartoon-anisotropic}.  
In isotropic colonies, neutral sector boundaries are perpendicular to the population edge.  
In contrast, sector boundaries in anisotropic colonies form a nontrivial angle~$\epsilon$ with the normal to the population edge, and this deviation depends on~$\theta$.  
The reflection symmetry~($x \to -x$) requires that~$\epsilon(0) = 0$, meaning vertical sector boundaries remain vertical, as seen in Fig.~\ref{fig:cartoon-anisotropic}.  
To leading order in~$\theta$, we approximate this deviation as~$\epsilon(\theta) \approx \epsilon'(0) \theta$.}  

\DS{This deflection of sector boundaries modifies the advection velocity in Eq.~(\ref{eqn:drift}), changing it from~$v( \theta)$ to~$v (\epsilon + \theta)$.  
Consequently,~$\beta$ takes the new value  
\[
\beta = v_0 (1 + \epsilon'(0)).
\]  
All other parameters in Eqs.~\eqref{eqn:FKPP} and \eqref{eqn:KPZ} remain unchanged compared to the isotropic case. } 

\DS{Since~$v_0$,~$\lambda$, and~$\beta$ are phenomenological parameters, they can be introduced without explicitly referencing their origins in~$v(\theta)$ and~$\epsilon(\theta)$.  
The key conclusion from the above analysis  is that for isotropic growth,~$v_0 = \lambda = \beta$, but for anisotropic growth, these three parameters can differ.  
Notably,~$v_0$ does not play a fundamental role in the dynamics, as it can be eliminated by shifting~$h \to h + v_0 t$.  
Thus, the ratio~$\beta / \lambda$ serves as a measure of anisotropy.}  

\DS{In general, anisotropic growth results in~$\beta / \lambda \neq 1$.  
However, this ratio remains equal to one if~$v_0 \epsilon'(0) = v''(0)$.  
This condition corresponds to trivial anisotropy that can be removed by rescaling~$x$ by~$v(\pi/2)$ and~$h$ by~$v(0)$—as is the case for the elliptical colonies in Fig.~\ref{fig:cartoon-anisotropic}.  
Note that such spatial rescalings can alter the values of~$\beta$ and~$\lambda$ but not their ratio.  
Thus, a model with~$\beta / \lambda \neq 1$ cannot be reduced to the isotropic case via simple rescaling.}  

\DS{As stated, Eqs.~\eqref{eqn:FKPP} and \eqref{eqn:KPZ} contain seven parameters:~$v_0$,~$\alpha$,~$\beta$,~$\lambda$,~$D_h$,~$D_f$, and~$s_0$.  
By shifting~$h \to h + v_0 t$, we eliminate~$v_0$.  
Additionally, three parameters (e.g.,~$s_0$,~$D_f$, and~$\beta$) can be set to unity via rescaling of~$h$,~$x$, and~$t$, and by possibly redefining~$f \to 1 - f$ through relabeling.  
This leaves three independent dimensionless parameters:  
\[
\frac{D_h}{D_f}, \quad \frac{\lambda}{\beta}, \quad \text{and} \quad \frac{\alpha \beta}{s_0 D_f}.
\]  }

\DS{In the following, we do not explicitly perform this nondimensionalization but instead analyze population dynamics in terms of intuitive variables:~$s_0$,~$\alpha$, and~$\beta$.  
However, we assume specific parameter signs to avoid redundant cases.  
Specifically, we take~$s_0 > 0$ by labeling the fitter type as type one, and we set~$\lambda > 0$ by flipping~$h \to -h$ if~$\lambda$ is negative.  
In the supplement~\cite{SI}, we discuss a lattice implementation of colony growth with~$\lambda < 0$ (see Figs.~S1 and S2).  
We also restrict our analysis to~$\beta > 0$, as the dynamics with~$\beta < 0$ differ significantly and warrant separate treatment.}  

\subsection{Gradient expansion}  

\DS{In addition to the geometric analysis presented above, Eqs.~\eqref{eqn:FKPP} and \eqref{eqn:KPZ} can be derived by employing general symmetry and locality principles in the limit of small gradients~$\partial f / \partial x$ and~$\partial h / \partial x$, as was done previously~\cite{drossel2000phase, horowitz2019bacterial, george_chirality_2018}.  
The reflection symmetry~$x \to -x$ forbids terms with an odd number of spatial derivatives, leaving only quadratic terms at the lowest order.  
However, not all such terms are included in Eqs.~\eqref{eqn:FKPP} and \eqref{eqn:KPZ}.}  

\DS{For instance, one could add a~$(\partial_x h)^2$ term to the FKPP equation to describe a fitness advantage of one type on sloped interfaces.  
Similarly, a~$\partial_x f \, \partial_x h$ term could be added to the KPZ equation to account for an increased expansion velocity in the presence of frequency gradients.  
Although these and similar terms are allowed by symmetry, they describe phenomena that have not yet been observed in a biological context.  
In the absence of empirical evidence supporting their contribution to colony dynamics, we neglect them in this study.}  

\DS{Finally, a gradient expansion does not impose constraints on the dependence of parameters on~$h$ and~$f$.  
Translational invariance rules out any~$h$-dependence, but in principle, all parameters could depend on~$f$.  
We have already discussed the possible dependence of~$\alpha$ on~$f$, which can account for species interactions affecting expansion velocity.  
Similarly, differences in motility patterns between the two types could lead to~$f$-dependent~$D_f$ and~$D_h$.  
The remaining parameters,~$\beta$ and~$\lambda$, could also vary with~$f$, for example, if the anisotropy axes of the two types are not aligned.  
Here, we focus on the simplest and perhaps most common case, where~$f$-dependence is negligible.  
We confirmed numerically that weak~$f$-dependence does not lead to qualitatively different results.  
However, a detailed investigation of how population dynamics are affected by~$f$-dependent parameters is beyond the scope of this study and warrants further exploration. } 

\subsection{A two-dimensional model of anisotropic growth}\label{sec:reaction_diffusion}  

\DS{While the primary approach of this work has been to characterize the dynamics governed by Eqs.~\eqref{eqn:FKPP} and \eqref{eqn:KPZ} with~$\lambda \neq \beta$, it is valuable to demonstrate that this analysis provides meaningful insights into the anisotropic growth of two-dimensional populations that are not constrained by the assumptions and simplifications leading to the coupled KPZ and FKPP equations.  
Reaction-diffusion models are widely used to simulate microbial growth \cite{korolev2012selective, mimura2000reaction, george_chirality_2018} and serve as a convenient framework to test the predictions of Eqs.~\eqref{eqn:FKPP} and \eqref{eqn:KPZ}, such as the new morphology described in the Results section. } 

\DS{There are several distinct ways to incorporate spatial anisotropy into a reaction-diffusion model~(see Supplemental Materials~\cite{SI}).  
Perhaps the simplest approach is to allow the diffusion coefficient to differ between the~$x$ and~$y$ directions.  
We introduced this generalization to a previously studied model of sector morphologies in isotropic growth~\cite{lee2022slow} leading to the following set of equations:  
\begin{align} \label{eqn:reaction_diffusion}
    \frac{\partial n_i(x,y,t)}{\partial t} &= 
    r_i(1-n)(n-K) \left(1 + \sum_j a_{ij} \frac{n_j}{n} \right) n_i \nonumber \\
    &\quad + D_{ix} \frac{\partial^2 n_i}{\partial x^2} 
    + D_{iy} \frac{\partial^2 n_i}{\partial y^2}.
\end{align}
\noindent Here, the indices~$i$ and~$j$ refer to specific species~(type one or type two), and $n = n_1 + n_2$ is the total density.  
The growth rate is specified by~$r_i$, and~$K$ is a constant, which is identical for both species.  
The matrix~$a_{ij}$ describes local competitive interactions between species, which do not affect their expansion velocities, and~$D_{ix}$ and~$D_{iy}$ are the diffusion coefficients.  
When~$D_{ix} = D_{iy}$, growth is isotropic and follows the description in Ref.~\cite{lee2022slow}.}  

\DS{When diffusion differs between the~$x$ and~$y$ directions, growth becomes anisotropic, but two distinct cases must be considered.  
For~$D_{1x} / D_{2x} = D_{1y} / D_{2y}$, the anisotropy can be removed by rescaling~$x \to x \sqrt{D_{1x} / D_{1y}}$ while keeping~$y$ unchanged.  
Since this form of anisotropy is rescalable, it does not lead to~$\lambda \neq \beta$ in the effective model~(see Supplementary Materials~\cite{SI}).  
However, rescalable anisotropy can still produce sector boundaries that are not perpendicular to the expansion front, as shown in Figs.~S4 and S5 of the Supplementary Materials.}  

\DS{A truly nontrivial anisotropy arises when~$D_{1x} / D_{2x} \neq D_{1y} / D_{2y}$, meaning the anisotropy axes are not aligned for the two species.  
This situation can occur, for example, if each species follows a distinct, misaligned nutrient gradient or if environmental factors impose species-specific anisotropy directions.  
Notably, this scenario is more general than the assumptions underlying Eqs.~\eqref{eqn:FKPP} and~\eqref{eqn:KPZ}, making the model in Eq.~(\ref{eqn:reaction_diffusion}) a more stringent test of our predictions.}  

\DS{In the following, we use Eq.~\eqref{eqn:reaction_diffusion} to illustrate anisotropic growth dynamics under various conditions and to confirm the existence of all sector morphologies predicted by Eqs.~\eqref{eqn:FKPP} and~\eqref{eqn:KPZ}.  
Additionally, in the Supplementary Materials~\cite{SI}, we examine a more general variant of Eq.~\eqref{eqn:reaction_diffusion} with potential density-dependent diffusion coefficients, as shown in Fig.~S3.}

\section{Results}
\subsection{Morphologies}  

\DS{The shape of a growing sector is the most salient prediction of Eqs.~\eqref{eqn:FKPP} and~\eqref{eqn:KPZ}.  
In previous work~\cite{swartz2023interplay}, we analyzed the isotropic limit~($\lambda = \beta$) and determined how~$\alpha$ and~$s_0$ influence the fate of a mutant.  
Both parameters describe the advantage of the mutant and can be affected by the same microscopic properties of the strains.  
For example, a higher growth rate~$r_i$ in Eq.~\eqref{eqn:reaction_diffusion} leads to larger~$\alpha$ and~$s_0$ in the effective model.  
However, this is not always the case.  
For instance, the parameters~$a_{ij}$ in the reaction-diffusion model affect~$s_0$ but not~$\alpha$.  
In general, certain biological mechanisms may affect~$\alpha$ and~$s_0$ similarly, while others may introduce a tradeoff between expansion rate and local competitiveness.  
The advantage of our phenomenological model is that we can treat~$\alpha$ and~$s_0$ as independent parameters to obtain general predictions.  
These predictions can then be used to understand more mechanistic models by imposing additional relationships between~$s_0$ and~$\alpha$ if such dependencies exist in the relevant biological context.}  

\DS{In the isotropic limit, Eqs.~\eqref{eqn:FKPP} and \eqref{eqn:KPZ} reproduce known sector morphologies, including a V-shaped dent, a composite bulge, and a circular arc.  
The V-shaped dent and circular arc have been observed experimentally~\cite{lee2022slow, korolev2012selective}, while the composite bulge was previously predicted using a geometric optics-inspired construction for light propagation~\cite{lee2022slow}.  
Our previous analysis in Ref.~\cite{swartz2023interplay} further clarified the origins of these morphologies and determined the invasion speed of the mutant as a function of microscopic parameters such as~$\alpha$ and~$s_0$.}  

\begin{figure*}
    \centering
    \includegraphics[width=\textwidth]{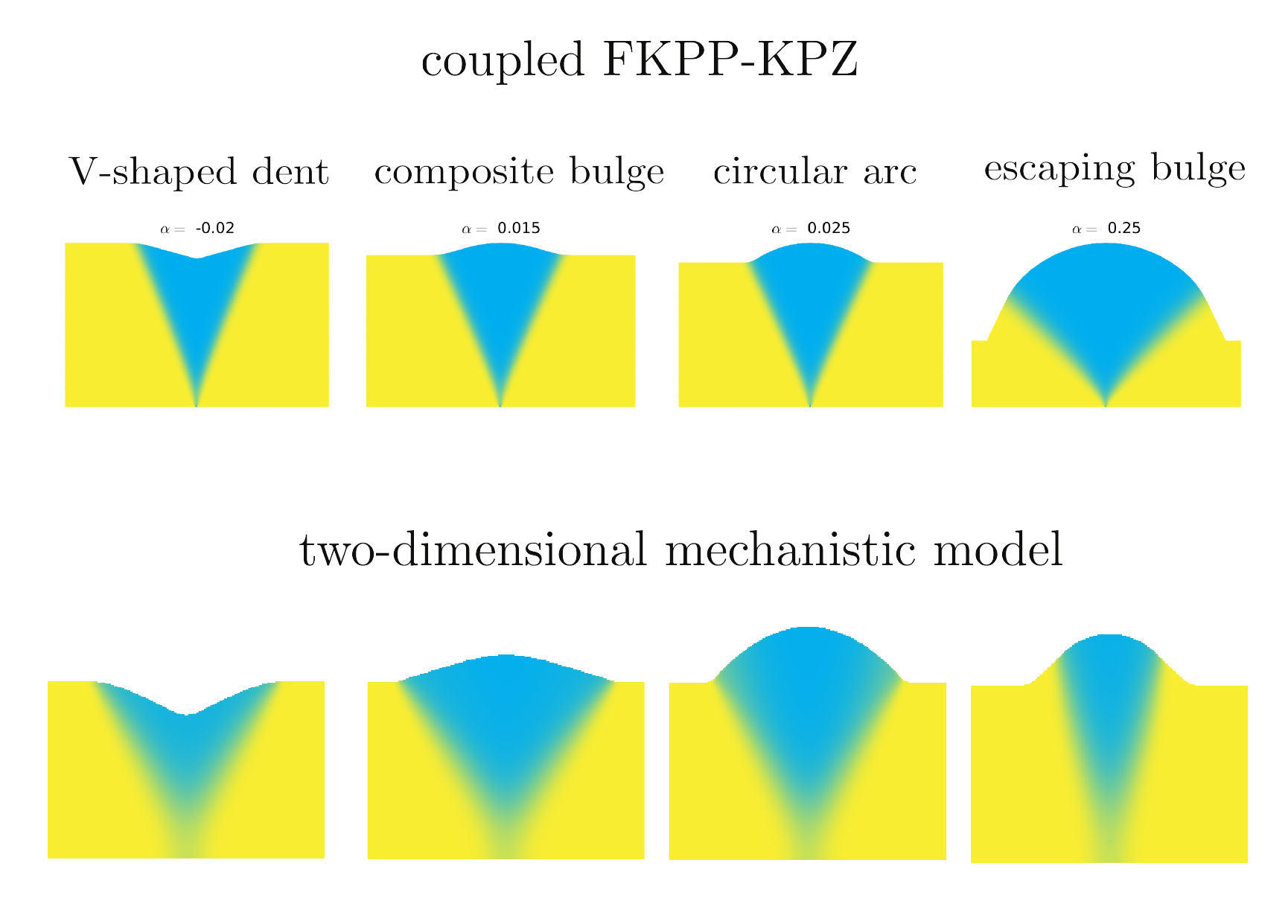}
    \caption{\textbf{Anisotropy enables an ``Escaping Bulge" sector morphology} (Color Online).  
    The top panel shows morphologies observed in numerical solutions of deterministic~\eqnS.  
    To the left~(negative~$\alpha$) is a V-shaped dent with straight edges.  
    Next, at small positive~$\alpha$, is the composite bulge morphology, characterized by constant limiting slopes surrounding a circular arc.  
    At larger~$\alpha$, the sector shape is a pure circular arc, commonly observed in experiments~\cite{korolev2012selective}.  
    The final shape is the newly identified \textit{escaping bulge}.  
    This morphology requires large~$\alpha$ but, more importantly, can only emerge in the presence of strong anisotropy~($\lambda \geq 2\beta$ for pulled waves).  
    The bottom panel demonstrates that the same morphologies exist in the anisotropic reaction-diffusion system described by Eq.~\eqref{eqn:reaction_diffusion}.  
    For the coupled FKPP-KPZ equations, the parameters used are $D_f = D_h = 1$, $v_0 = 0.15$, $\lambda = 20$, $s_0 = 0.25$, and $\beta = 5$.  
    For the two-dimensional mechanistic model, all panels use $r_1 = r_2 = 1$, $K = 0.1$, $D_{1x} = D_{1y} = D_{2y} = 0.02$, and $D_{2x} = 0.09$ (with type 1 being blue).  
    The interaction strengths $a_{ij}$ used for the different morphologies are, from left to right,  
    $(a_{11}, a_{21}, a_{12}, a_{22}) = (-0.7, -0.7, 0.7, 0), (0.4, 0.4, 0.9, 0), (1, 1, 0.6, 0), (1, 1, 0, 0)$.  
    }
    \label{fig:morphologies}
\end{figure*}

\DS{Here, we generalize this analysis to anisotropic growth and investigate how morphologies and invasion velocities are affected by~$\beta / \lambda$.  
The main results are summarized in Fig.~\ref{fig:morphologies}, which illustrates how sector morphologies change with the expansion rate difference~$\alpha$ between the blue and yellow strains.  
As~$\alpha$ increases from negative to positive values, the morphology transitions from a \textit{V-shaped dent} to a \textit{composite bulge} and then to a \textit{circular arc}.  
These transitions occur in both isotropic and anisotropic growth, but above a critical ratio of~$\beta / \lambda$, a qualitatively new morphology emerges when~$\alpha$ is sufficiently large; see the last panel in Fig.~\ref{fig:morphologies}.  
This new morphology, which we term an \textit{escaping bulge}, differs fundamentally from the other three sector types.  
In this case, the boundary between the blue and yellow strains is preceded by a deformation in the colony shape deep within the region occupied by the less fit strain.}  

\DS{In the following, we fully characterize this new morphology and determine the invasion speeds of both the strain boundary and the deformation.  
We also analyze the effects of anisotropy on the invasion velocity for the V-shaped dent, composite bulge, and circular arc.}  

\subsection{Invasion speed: Pulled Waves}  

\begin{figure}
    \centering
    \includegraphics[width=0.5\textwidth]{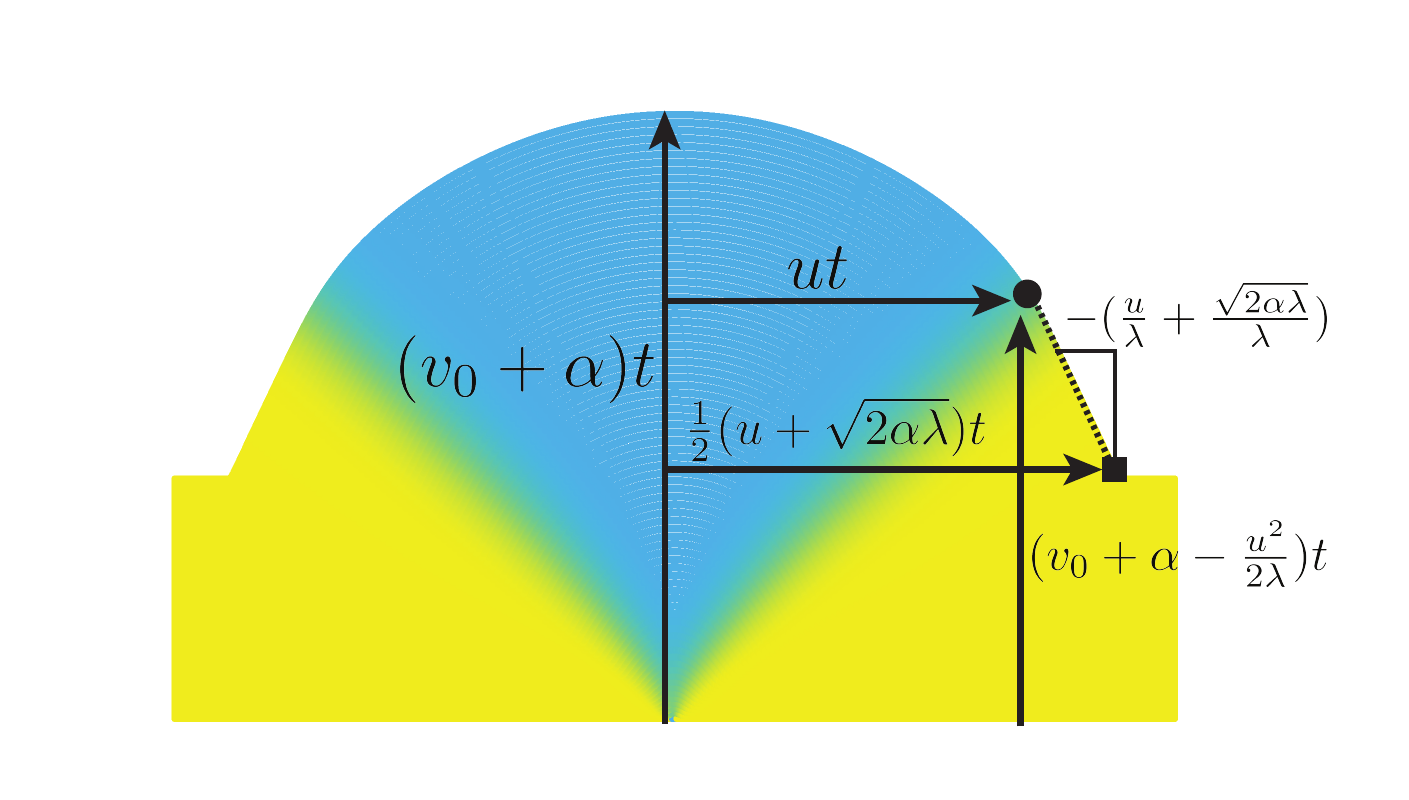}
    \caption{\textbf{Schematic of the escaping bulge morphology.}  
    This diagram illustrates the key geometric features involved in deriving the invasion speed for the escaping bulge morphology~(Eq.~\eqref{eqn:u_escaping}).  
    The solid dot marks the instantaneous location of the sector boundary at time~$t$, while the square indicates the location of the shock in the invaded strain's morphology, which moves with speed~$c$ as given in Eq.~\eqref{eqn:shock_speed}.  
    The dotted line represents the limiting slope of the escaping bulge far ahead of the invasion front, as described by Eq.~\eqref{eqn:sigma_escaping}.  
    }
    \label{fig:escaping_schematic}
\end{figure}

The key observable predicted by \eqnS is the invasion velocity~$u$, i.e., the velocity at which one strain (blue) displaces the other strain (yellow) along the $x$-axis, see Fig.~\ref{fig:escaping_schematic}.  
By convention, we take~$u$ to be positive for a right-moving wave.  
In this section, we compute~$u$ from a traveling-wave solution to \eqnS for all morphologies.  
Notably, the FKPP equation~\eqref{eqn:FKPP} describes a special class of traveling front solutions known as~\textit{pulled waves}, which are dominated by the dynamics at the leading edge where~$f$ is small~\cite{van2003front}.  
\DS{The case of~\textit{pushed waves}, which arises when~$s_0$ is not constant but strongly increases with~$f$, is analyzed in the next section.}  

We begin by recalling two relevant velocities in Eqs.~\eqref{eqn:FKPP} and \eqref{eqn:KPZ}.  
The first is the Fisher speed,~$u_0= 2\sqrt{s_0 D_f}$, which is the speed of a pulled wave in the absence of coupling to~$h$~\cite{kolmogorov1937moscow, fisher1937wave}.  
Apart from the factor of~$2$, this velocity follows from dimensional analysis of the FKPP equation.  
The second velocity, which can also be obtained by dimensional analysis, is the natural speed in the KPZ equation.  
It describes invasion in the circular arc morphology~\footnote{The KPZ equation is a first-order approximation to circular colony growth. The height field emerging from KPZ is parabolic:~$h(x,t) \approx (v_0 + \alpha)t - \frac{x^2}{2\lambda t}$.}  
and is given by~$u_{\text{kpz}} = \sqrt{2\alpha \lambda}$; see Ref.~\cite{swartz2023interplay}.  

\DS{Previously, we demonstrated that the invasion speed of an isotropic pulled wave is given by the larger of these two velocities~\cite{swartz2023interplay}.  
For small~$\alpha$, the invasion speed is determined entirely by the FKPP equation, and the coupling term has a negligible effect.  
For large~$\alpha$, the situation is reversed: the KPZ equation~(with~$f$ set to~$1$) controls the invasion speed.}  

\begin{figure}
    \centering
    \includegraphics[width = 0.5\textwidth]{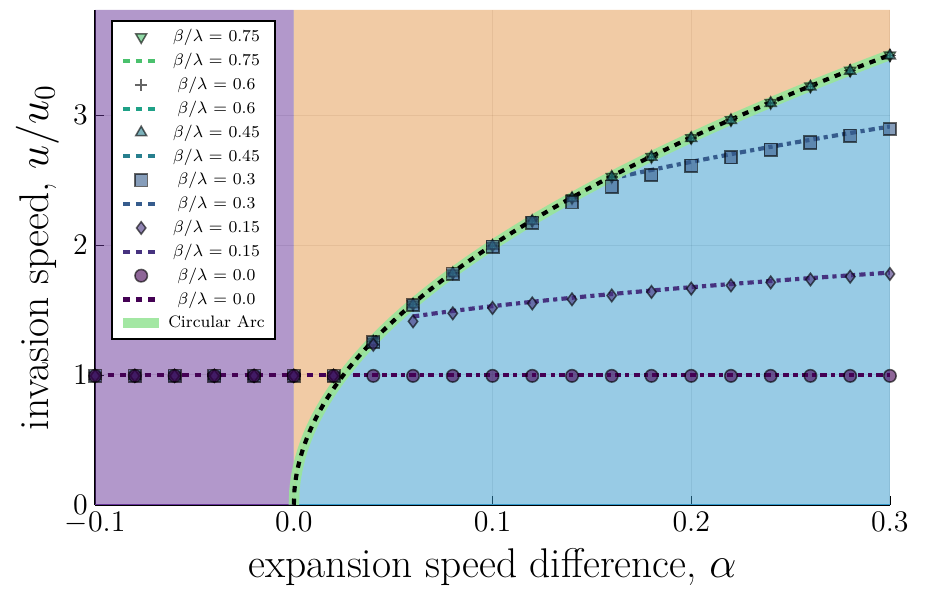}
    \caption{\textbf{Three regimes for anisotropic pulled  waves} (Color Online). The mutant can invade with the Fisher velocity $u_0$, the velocity of the circular arc~$u_{\text{kpz}} = \sqrt{2\alpha \lambda}$~(dashed black line with green ribbon), or the escaping bulge velocity predicted by Eq.~\eqref{eqn:u_escaping}~(dashed lines of different colors) depending on the expansion speed difference~$\alpha$ and the degree of anisotropy~$\beta/\lambda$. The markers are obtained from numerical solutions of \eqnS. Regions of the plot are colored based on the morphology, matching the phase diagram in Fig.~\ref{fig:pulled_phase}. When $\alpha < 0$ the morphology is a V-shaped dent (purple color). When $\alpha > 0$ and $u > u_{\text{kpz}}$ the morphology is composite (orange color). When $\alpha > 0$ and $u = u_{\text{kpz}}$ the morphology is a circular arc (green region with black dashed center line). When $\alpha > 0$ and $u < u_{\text{kpz}}$ we find the escaping bulge morphology (blue color). Parameters are: $v_0 = 1$, $\lambda = 20$, $D_h = 1$, $D_f = 1$, and~$s_0 = 0.25$.
    } 
    \label{fig:uvsalpha_pulled}
\end{figure}

\DS{Since neither the Fisher speed $u_0$ nor the circular arc velocity $u_{\text{kpz}}$ depends on the advective coupling~$\beta$, it is natural to expect that anisotropy, which enters the model only through $\beta$, should neither affect these velocities nor the transition between the two regimes.  
Indeed, numerical simulations confirm this behavior (see Fig.~\ref{fig:uvsalpha_pulled}).  
Moreover, all relevant calculations in Ref.~\cite{swartz2023interplay} do not depend on~$\beta$ and thus remain valid in the anisotropic case considered here.}
For example, in the isotropic model the value of $\alpha$ marking the transition between the composite bulge and arc morphologies is obtained by equating the Fisher speed $u_0$ to the circular arc speed $u_\text{kpz}$. The Fisher speed $u_0$ is found by linearizing Eq.~\eqref{eqn:FKPP} about the leading edge of the invasion front. In this regime, the advective coupling term $\beta\,\partial_x h\,\partial_x f$ is subleading and does not provide a correction to the Fisher speed $u_0$. The circular arc speed $u_{\text{kpz}}$, in contrast, arises solely from the KPZ equation with $f \approx 1$ and so cannot depend on $\beta$ either. 

However, the escaping bulge morphology does not occur in the isotropic case and requires separate analysis. Our calculations below demonstrate that the escaping bulge is only stable when~$\beta / \lambda$ is sufficiently small.  
For weak anisotropy or when~$\beta > \lambda$, we observe the same dynamics as in the isotropic case—namely, the transition from the Fisher speed~$u_0$ to the circular arc velocity~$u_\text{kpz}$ as~$\alpha$ increases.  
For small~$\beta$, a second transition occurs as~$\alpha$ increases further: the invasion velocity continues to increase with~$\alpha$, but at a slower rate than predicted by the KPZ velocity, and the circular arc morphology is replaced by an escaping bulge.  

\DS{In the remainder of this section, we compute both the invasion speed of the escaping bulge morphology and the critical anisotropy strength~$\beta / \lambda$ necessary for its emergence.  To carry out this computation, we switch to a reference frame co-moving with the sector boundary in both horizontal and vertical directions; see Fig.~\ref{fig:escaping_schematic}.  
The horizontal velocity~$u$ can be eliminated via a change of variables:~$z = x - u t$, with~$u$ still to be determined.} 

\DS{The vertical position of the sector boundary can be determined as follows.  
The inner region of the escaping bulge contains a single strain and should, therefore, follow a self-similar solution of the KPZ equation—either a line or a parabola~\cite{swartz2023interplay}.  
Thus, the inner blue region in Fig.~\ref{fig:escaping_schematic} is described by  
\[
h(x,t) \approx (v_0 + \alpha) t - \frac{x^2}{2 \lambda t}\,.
\]  
This solution extends until the sector boundary is reached at~$x = \pm u t$.  
Eliminating~$x$ from these two equations, we find that the boundary is located at  
\[
h = \left(v_0 + \alpha - \frac{u^2}{2\lambda} \right) t\,;
\]  
see the rightmost vertical arrow in Fig.~\ref{fig:escaping_schematic}.  
Hence, we redefine~$h$ as~$h + (v_0 + \alpha - u^2 / 2\lambda)t$.}  

\DS{Upon subtracting both the horizontal and vertical positions of the sector boundary from~$x$ and~$h$, we obtain}  
\begin{align}
    - u f' &= s_0 f (1-f) + D_f f'' + \beta f' h' \, ,\label{eqn:fwave_pulled} \\
    -u h' &= \frac{u^2}{2\lambda} + \alpha (f-1) + \frac{\lambda}{2} (h')^2 + D_h h'', \label{eqn:hwave_escaping_pulled} 
\end{align}  
\noindent \DS{where derivatives are taken with respect to the co-moving coordinate~$z$.  
The invasion velocity~$u$ is then determined by seeking a nontrivial stationary solution to the equations above.}  

\DS{Since the velocity of pulled waves is determined by the region where~$f$ vanishes, we analyze the behavior of~$h$ at~$z \to +\infty$.  
From Eq.~\eqref{eqn:hwave_escaping_pulled}, it immediately follows that}  
\begin{equation}
    h'(z\to +\infty) = -\frac{u}{\lambda} - \frac{\sqrt{2\alpha \lambda}}{\lambda}\,,
    \label{eqn:sigma_escaping}
\end{equation}
\noindent \DS{because~$f(z\to +\infty) = 0$ and~$h''$ vanishes for a constant slope.  
The limiting slope given by Eq.~\eqref{eqn:sigma_escaping} is shown as the dashed line in Fig.~\ref{fig:escaping_schematic}.  
The second transition to a flat front occurs in a region occupied only by the yellow strain and does not affect the calculation above} \footnote{Greater care must be taken for negative~$\beta$, see Ref.~\cite{lam2022asymptotic}.}.  

\DS{Equation~\eqref{eqn:fwave_pulled} then reduces to a standard FKPP equation in a co-moving reference frame with an additional drift velocity~$-\beta h'(z\to +\infty)$ given by Eq.~\eqref{eqn:sigma_escaping}.  
Thus,}  
\[
u=2\sqrt{s_0 D_f}+\beta\left(\frac{u}{\lambda} + \frac{\sqrt{2\alpha \lambda}}{\lambda} \right),
\]  
\DS{which simplifies to}  
\begin{equation}
    u = \frac{2\sqrt{s_0 D_f} + \frac{\beta}{\lambda} \sqrt{2\alpha\lambda}}{1-\beta/\lambda}\,.  
    \label{eqn:u_escaping}
\end{equation}  
\DS{This solution agrees with numerical simulations} \footnote{All code used for simulations is available on GitHub~\cite{aniso_final_repo}.}  
\DS{for appropriate parameter values, as shown in Fig.~\ref{fig:uvsalpha_pulled}.  
However, it diverges for~$\beta = \lambda$ and, in general, is valid only in the regime where the escaping bulge is observed.}  

\DS{We determine the region of applicability by considering the vertical velocity of the sector boundary,  
which we previously found to be~$v_0 + \alpha - u^2/(2\lambda)$.  
For the escaping bulge to persist at long times, this velocity must exceed the upward velocity of the flat front,~$v_0$.  
Thus, an escaping bulge is only possible when}  
\[
u < \sqrt{2\alpha \lambda}, \quad \text{i.e.,} \quad u < u_{\text{kpz}}.  
\]  
\DS{From Eq.~\eqref{eqn:u_escaping}, it follows that the escaping bulge occurs only for}  
\[
\beta < \lambda/2.
\]  
\DS{In other words, the degree of anisotropy must exceed a critical value for the escaping bulge to emerge.  
Figure~\ref{fig:pulled_phase} illustrates the transitions between all possible regimes.  
In the Supplementary Materials~\cite{SI}, we provide example morphologies for various values of~$\alpha$ and~$\beta$ (see Fig.~S7)  
and demonstrate that small~$\beta$ is required to observe the escaping bulge morphology.}  

\begin{figure}
    \centering
    \includegraphics[width=0.5\textwidth]{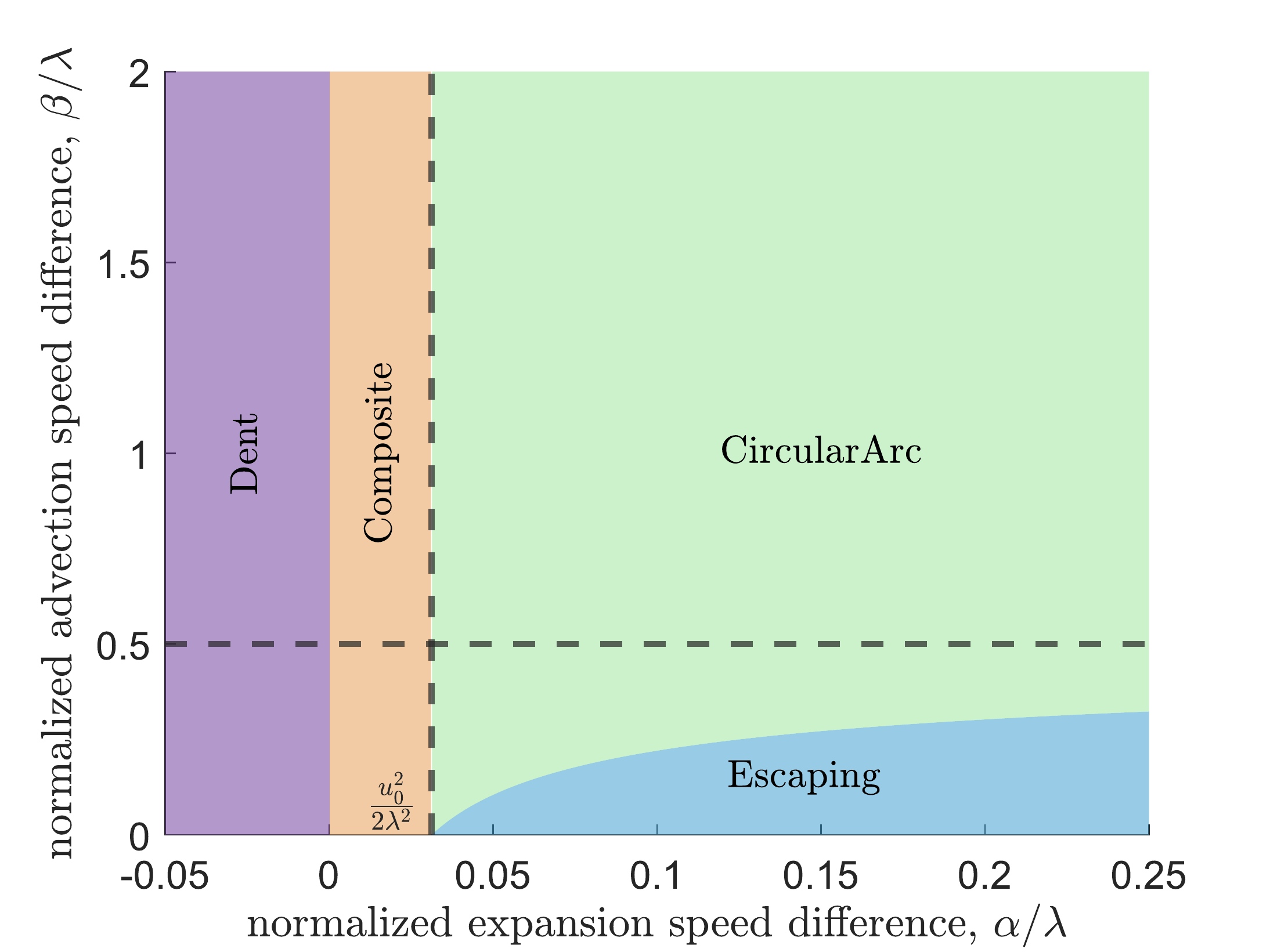}
    \caption{\textbf{Predicted phase diagram of morphologies for pulled waves based on~\eqnS} (Color Online).  
    The isotropic case corresponds to the line~$\beta / \lambda = 1$.  
    Small anisotropy does not produce new morphologies, whereas the escaping bulge can only exist for~$\beta / \lambda < 1/2$ (horizontal dashed line).  
    Parameters not shown are~$u_0 = 5$ and~$\lambda = 20$.  
    }
    \label{fig:pulled_phase}
\end{figure}

\begin{figure}
    \centering
    \includegraphics[width=0.5\textwidth]{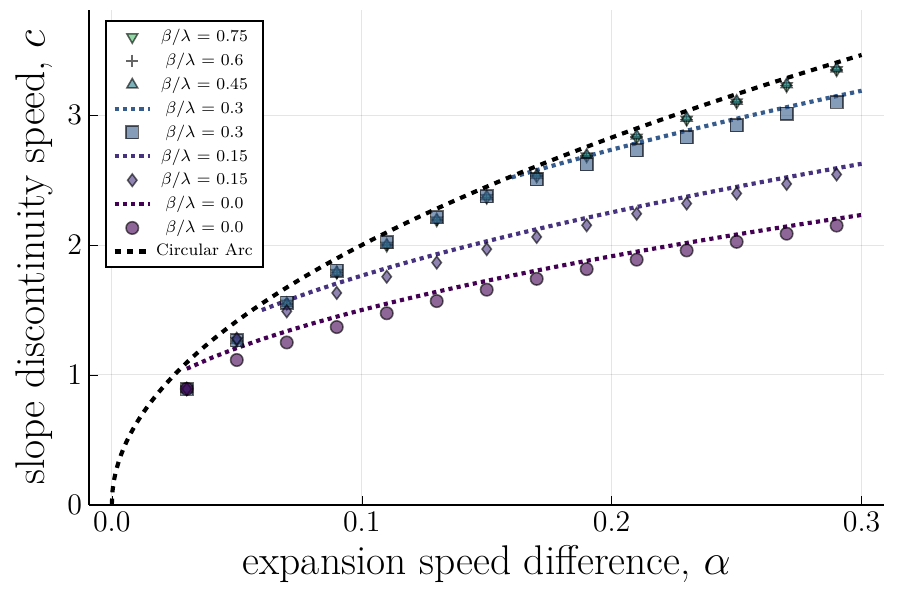}
    \caption{\textbf{Speed of the discontinuity in the slope.}  
    The shock speed is measured by tracking the point where~$h = 1.01 v_0 t$,  
    i.e., where the front is~$1\%$ higher than the flat front.  
    The dashed black line represents the invasion speed of the circular arc morphology,~$u_{\text{kpz}} = \sqrt{2 \alpha \lambda}$,  
    while the dashed colored lines correspond to the analytical predictions from Eqs.~\eqref{eqn:u_escaping} and~\eqref{eqn:shock_speed}.  
    The parameters are the same as in Fig.~\ref{fig:uvsalpha_pulled}.  
    }
    \label{fig:shock_speed}
\end{figure}

Another key feature of the escaping bulge morphology is a shock-like singularity ahead of the sector boundary,  
depicted as the square marker in Fig.~\ref{fig:escaping_schematic}.  
The speed of this singularity,~$c$, is greater than~$u$ and can be determined by finding the intersection point  
between the unperturbed front of the blue strain—moving upward with velocity~$v_0$—and the straight slope extending downward from the sector boundary.  
Since we have already determined both the value of this slope and the location of the sector boundary, we find that  
\begin{equation}
    c = \frac{1}{2}(u + \sqrt{2 \alpha \lambda}).
    \label{eqn:shock_speed}
\end{equation}  
This result is confirmed by simulations in Fig.~\ref{fig:shock_speed}, further supporting the conclusion that  
the escaping bulge can only occur when~$u < u_{\text{kpz}}$.  

\subsection{Invasion speed: Pushed Waves}
The FKPP equation considered so far is a special case of a broader class of equations where~$s_0$ is replaced by an arbitrary function~$s(f)$.  
This generalization accounts for frequency-dependent selection, which often arises due to metabolite exchanges or other forms of collective or cooperative growth strategies~\cite{swartz2023interplay, gandhi2016range, menon2015public, lavrentovich2014asymmetric, nadell2016spatial, korolev2011competition}.  
Although our analysis applies to any choice of~$s(f)$, we restrict our discussion to~$s(f) = s_0 (f - f_0)$ with~$s_0 > 0$.  
\DS{This form, depending on the value of~$f_0$, captures different types of invasion dynamics (pushed and pulled traveling waves) while introducing minimal additional complexity~\cite{swartz2023interplay, birzu_fluctuations_2018, roques2012allee}. } 

\DS{Furthermore, we take advantage of exact analytical solutions for the generalized FKPP equation with this linear form of~$s(f)$.  
Thus, we now replace Eq.~\eqref{eqn:FKPP} with  
\begin{equation}
    \frac{\partial f}{\partial t} = s_0 (f-f_0) f (1-f) + D_f \frac{\partial^2 f}{\partial x^2} + \beta \frac{\partial h}{\partial x} \frac{\partial f}{\partial x}\,,
    \label{eqn:fpushed}
\end{equation}
\noindent while keeping Eq.~\eqref{eqn:KPZ} unchanged. } 

For~$f_0 < -1/2$, the solution of the generalized FKPP equation~(without coupling to~$h$) corresponds to a pulled wave.  
In this case, all of our previous results remain valid if we replace~$s_0$ with~$-s_0 f_0$, which represents the effective competitive advantage in the limit of small~$f$ in this model~\cite{birzu_fluctuations_2018}.  
For~$f_0 > -1/2$, however, the traveling-wave solutions become pushed, meaning that their velocity depends not only on~$s(0)$,  
which governs dynamics at the invasion edge, but also on the entire shape of~$s(f)$, since competition across the entire invasion front influences the invasion velocity~\cite{van2003front, birzu_fluctuations_2018, meerson2011velocity, paquette1994structural, rocco2001diffusion, mikhailov1983stochastic, roques2012allee}.  

Within the class of pushed waves, one typically distinguishes between waves propagating into a metastable state~($f_0 > 0$)  
and those propagating into an unstable state~($f_0 < 0$).  
In our previous work~\cite{swartz2023interplay}, we showed that these two subclasses behave differently at large negative~$\alpha$.  
In the former case, the wave undergoes \textit{reversal}, with~$u$ becoming negative, while in the latter case, the wave effectively behaves as a pulled wave,  
and the invasion velocity approaches~$2\sqrt{-s_0 f_0 D_f}$.  

The behavior at negative~$\alpha$ primarily affects the V-shaped dent morphology, which appears in both isotropic and anisotropic growth.  
The only distinction between these cases is a quantitative shift in invasion speed.  
Since anisotropic growth does not introduce qualitative changes for~$\alpha < 0$,  
the distinction between metastable and unstable invasion at large negative~$\alpha$ should remain unchanged from our previous work.  
Here, we instead focus on the case of positive or slightly negative~$\alpha$,  
where anisotropy introduces nontrivial effects, which can be explored perturbatively. In previous work~\cite{swartz2023interplay}, we demonstrated that the magnitude of any slope in the height function $h$ is set by the expansion speed difference $\alpha$. This motivates a small $\alpha$ expansion where terms proportional to $\partial_x h$ are themselves treated as small perturbations.

To determine~$u$, we treat the coupling between the KPZ and FKPP equations as a perturbation.  
We begin with the unperturbed solution of Eq.~\eqref{eqn:fpushed}, i.e., with~$\beta$ set to zero.  
In general, these solutions must be obtained numerically, but for the linear~$s(f)$ case, the profile shape~$f^{(0)}(z)$ and invasion speed~$u^{(0)}$ are known exactly~\cite{fife1977approach}:  
\begin{align}
    f^{(0)} &= \frac{1}{1+e^{\sqrt{\frac{s_0}{2D_f}} z}} \,,\\
    u^{(0)} &= \sqrt{\frac{s_0 D_f}{2}} (1-2f_0\,).
\end{align}  
This zeroth-order solution can then be used to determine a correction to the colony shape~$h$ from Eq.~(\ref{eqn:KPZ}).  

For the V-shaped dent and composite bulge, this procedure was carried out in Ref.~\cite{swartz2023interplay}.  
Here, we focus instead on the new escaping bulge morphology.  
As in our analysis of pulled waves, we first switch to a reference frame co-moving with the sector boundary in both horizontal and vertical directions; see Fig.~\ref{fig:escaping_schematic}.  

\DS{The result is identical to Eq.~\eqref{eqn:hwave_escaping_pulled},  
but now with a pushed rather than a pulled zeroth-order solution for the genetic wave~$f^{(0)}$:}  
\begin{equation}
    -u h'(z) = \frac{u^2}{2\lambda} + \alpha (f^{(0)} - 1) + \frac{\lambda}{2} (h')^2 + D_h h''.  
    \label{eqn:hwave_escaping} 
\end{equation}  
While Eq.~\eqref{eqn:hwave_escaping} can be reduced to a Riccati equation,  
it is simpler and perhaps more informative to analyze the colony shape~$h$ in the geometric optics limit by setting~$D_h = 0$~\cite{korolev2012selective, lee2022slow}.  
\DS{In this limit, we allow for rapid changes in the slope~$h'(z)$ that would otherwise be smoothed by the diffusion term in the KPZ equation.  
Our previous work showed that this approximation provides qualitatively correct behavior even for large values of~$D_h$.  
However, for an accurate quantitative prediction, Eq.~(\ref{eqn:hwave_escaping}) must be solved numerically.  
The approximate result reads}  
\begin{equation}
    h'(z) = -\frac{u}{\lambda} - \sqrt{\frac{2\alpha}{\lambda} (1 - f^{(0)}(z))},
    \label{eqn:hprime_escaping}
\end{equation}  
\noindent \DS{which allows us to evaluate the coupling term in the FKPP equation.  
Rewriting the FKPP equation in terms of~$f$, we obtain}  
\begin{equation}\label{eqn:schematic_perturbation}
    -u f'(z) = R(f) + D_f f''(z) + P(z),
\end{equation}  
\DS{where~$P(z)$ is a~$z$-dependent perturbation arising from the coupling term, given by}  
\[
P(z) = \beta h'(z) f^{(0)\prime}(z).
\]  
\DS{Here,~$R(f)$ represents the original selection term with~$s(f)$ for brevity.  }

\DS{
The perturbative solution to Eq.~\eqref{eqn:schematic_perturbation} can be obtained for $P$ using standard methods~\cite{ birzu_fluctuations_2018, meerson2011velocity, paquette1994structural, rocco2001diffusion, mikhailov1983stochastic} with the following result
\begin{equation}
    u = u^{(0)} - \frac{\int_{-\infty}^{\infty} P(z) f^{(0)\prime} e^{u^{(0)} z / D_f}dz}{\int_{-\infty}^{\infty} (f^{(0)\prime})^2 e^{u^{(0)} z / D_f}dz}\,.
    \label{eqn:old_perturbation}
\end{equation}
\noindent Upon using Eq.~(\ref{eqn:old_perturbation}) and the explicit form of~$P$, we find that
\begin{equation}
    u = u^{(0)} - \beta \frac{\int_{-\infty}^{\infty} h^{\prime} (f^{(0)\prime})^2 e^{u^{(0)} z / D_f}dz}{\int_{-\infty}^{\infty} (f^{(0)\prime})^2 e^{u^{(0)} z / D_f}dz}\,.
    \label{eqn:new_perturbation}
\end{equation}
}

\begin{figure}
    \centering
    \includegraphics[width = 0.5\textwidth]{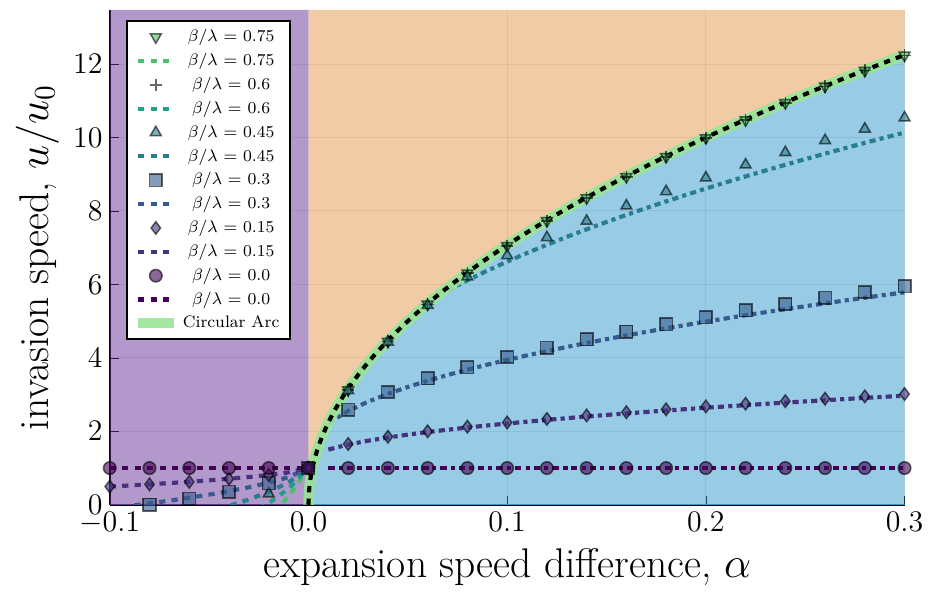}
    \caption{\textbf{Invasion speed for pushed waves} (Color Online).  
    Symbols represent numerical solutions of Eqs.~\eqref{eqn:KPZ} and \eqref{eqn:fpushed},  
    while dashed colored lines indicate theoretical predictions.  
    Three theoretical solutions are shown:  
    (i) The perturbative solution for a composite bulge and V-shaped dent, valid for~$\alpha < 0$ and slightly positive~$\alpha$,  
    given by Eqs.~\eqref{eqn:u_pert_old} and \eqref{eqn:sig_pert_old} (where~$\kappa$ was evaluated numerically as~$\kappa \approx 0.51$).  
    (ii) The circular arc solution, where~$u = u_{\mathrm{\textsc{kpz}}} = \sqrt{2\alpha \lambda}$, shown as a dashed black line with a green ribbon.  
    (iii) The escaping bulge solution, given by Eq.~\eqref{eqn:u_pushed}, shown as a dashed line for larger positive~$\alpha$  
    (with~$\gamma = 0.83$ obtained from Eqs.~\eqref{eqn:hprime_escaping} and \eqref{eqn:gamma}).  
    Regions are color-coded as in Fig.~\ref{fig:pulled_phase}.    
    Parameters:~$v_0 = 1$, $\lambda = 20$, $D_h = 1$, $D_f = 1$, $s_0 = 1$, and~$f_0 = 0.1$.  
    }
    \label{fig:uvsalpha_pushed}
\end{figure}

The final result for~$u$ can then be obtained by using the approximate solution for~$h'$ from Eq.~\eqref{eqn:hprime_escaping}:
\begin{equation}
    u = \frac{u^{(0)} + \gamma \frac{\beta}{\lambda} \sqrt{2\alpha\lambda}}{1-\frac{\beta}{\lambda}},\
    \label{eqn:u_pushed} 
\end{equation}
\noindent where
\begin{equation}
    \gamma = \frac{\int_{-\infty}^{\infty} \sqrt{1-f^{(0)}}(f^{(0)\prime})^2 e^{u^{(0)} z / D_f}dz}{\int_{-\infty}^{\infty} (f^{(0)\prime})^2 e^{u^{(0)} z / D_f}dz}\,,
    \label{eqn:gamma}
\end{equation}
\noindent is a numerical factor, which depends on~$f_0$ and can in principle be evaluated analytically. Note that only the equation for~$\gamma$ is affected by our approximation that~$D_h=0$ while the equation for~$u$ is valid for all $D_h\ge0$. These analytical results match well with the numerical results shown in Fig.~\ref{fig:uvsalpha_pushed} even though we used~$D_h=1$ in simulations.

For~$f_0 \to -1/2$, i.e. on the boundary with pulled waves,~$\gamma\to1$ in agreement with our results in the previous section. For larger values of~$f_0$, this prefactor is strictly less than one, and~$\gamma\to 24/35$ as~$f_0\to1/2$, which is the maximal value of~$f_0$ consistent with our choice of positive~$u^{(0)}$.

\begin{figure}
    \centering
    \includegraphics[width=0.5\textwidth]{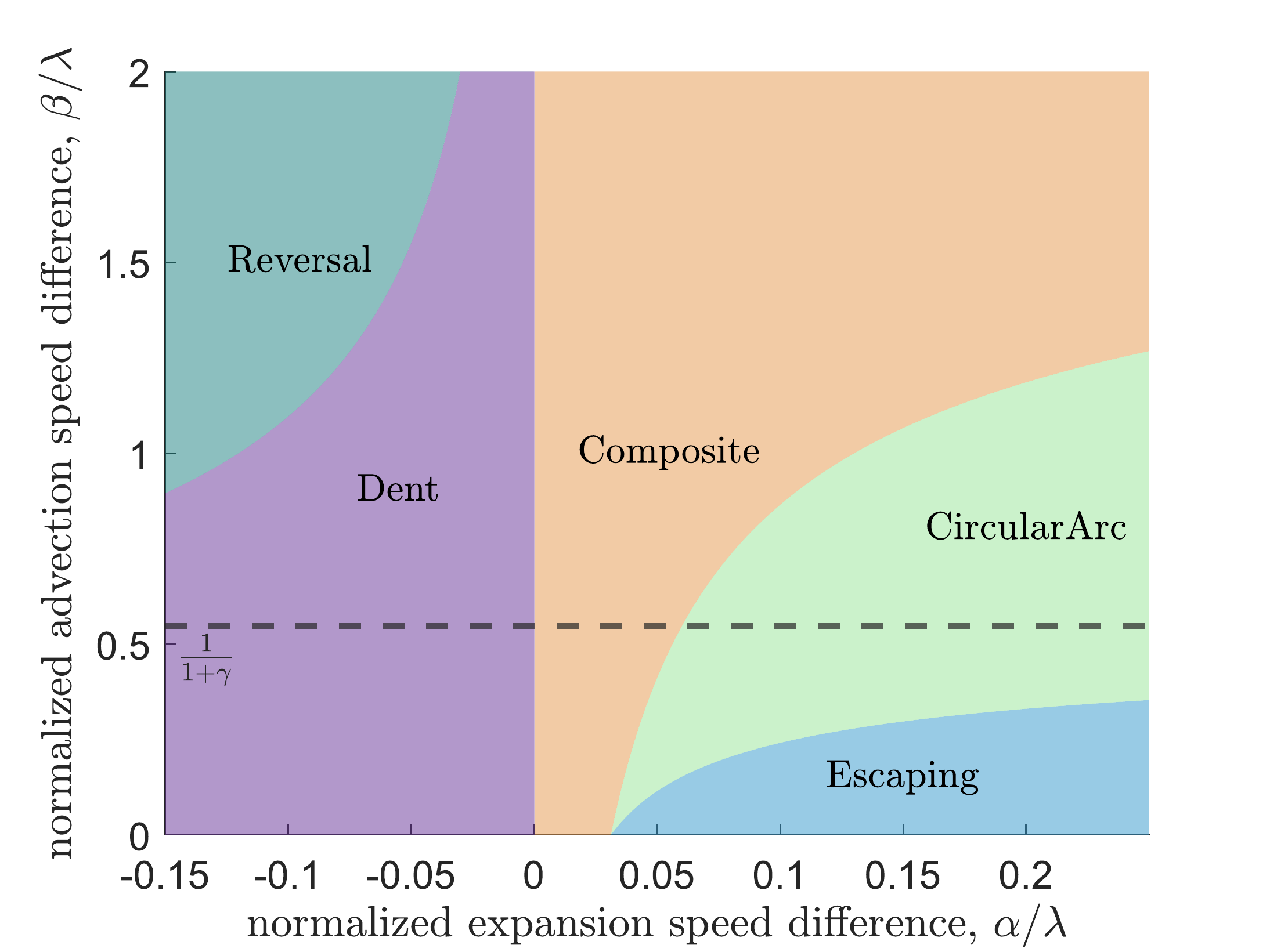}
    \caption{\textbf{Predicted phase diagram of morphologies for pushed waves} (Color Online).  
    Phase boundaries are determined by matching the invasion speeds computed for the three morphologies  
    as a function of the normalized expansion speed difference~$\alpha / \lambda$ and the normalized anisotropy~$\beta / \lambda$.  
    In particular, the boundary between the composite bulge and circular arc is obtained  
    by equating the natural KPZ velocity~$\sqrt{2\alpha\lambda}$ to~$u$ from Eq.~\eqref{eqn:old_perturbation},  
    as was previously done for isotropic growth~\cite{swartz2023interplay}.  
    The boundary between the circular arc and escaping bulge morphologies is given by Eq.~\eqref{eqn:u_pushed}.  
    Notably, the escaping bulge morphology is accessible only for~$\beta < \lambda / (1 + \gamma)$.  
    For~$f_0 > 0$, pushed waves also exhibit a \textit{Reversal} phase,  
    where the direction of invasion flips due to an overwhelming growth rate disadvantage~\cite{swartz2023interplay}.  
    Parameters:~$u^{(0)} = 5$, $\lambda = 20$, $\kappa = 0.51$, and~$\gamma = 0.83$, corresponding to~$f_0 = 0.1$.  
    }
    \label{fig:pushed_phase}
\end{figure}

\DS{The region of parameter space that results in an escaping bulge can be determined similarly to the case of pulled waves.  
Specifically, by setting the vertical velocity of the sector boundary to~$v_0$,  
we find that an escaping bulge occurs only when~$u < u_{\text{kpz}}$.  
From Eq.~(\ref{eqn:u_pushed}), this condition can be satisfied only if  
\[
\beta < \frac{\lambda}{1+\gamma}.
\]  
Since~$\gamma$ is less than one for pushed waves,  
we conclude that an escaping bulge requires weaker anisotropy in pushed waves compared to pulled waves.  
The transitions between different morphologies for pushed waves with positive~$f_0$ are summarized in Fig.~\ref{fig:pushed_phase}.}

\DS{We also note that all calculations leading to Eq.~\eqref{eqn:shock_speed} are purely geometric  
and do not depend on whether the wave is pushed or pulled.  
Thus, the result for the speed of the shock-like deformation of the front holds for pushed waves as well.}  

\DS{Finally, we outline the effect of anisotropy on~$u$ for morphologies other than the escaping bulge.  
In Ref.~\cite{swartz2023interplay}, we derived an equation identical to Eq.~(\ref{eqn:new_perturbation}),  
except it contained~$\lambda$ instead of~$\beta$, since these parameters are equal in the isotropic case.  
Thus, our previous results can be generalized to anisotropic growth by replacing~$\lambda$ with~$\beta$ where appropriate.  
However, for the V-shaped dent and composite bulge morphologies, one must use the expression for~$h'$  
derived in Ref.~\cite{swartz2023interplay} rather than Eq.~(\ref{eqn:hwave_escaping}),  
which was specifically derived for the escaping bulge morphology.  
The velocity of the circular arc morphology, given by~$u_{\text{kpz}}$, does not depend on~$\beta$ and remains unaffected by anisotropy.}  

\DS{The invasion velocity for the composite bulge or the V-dented front is given by  
\begin{equation}\label{eqn:u_pert_old}
    u = u^{(0)} - \kappa \frac{\beta}{\lambda} \sigma,
\end{equation}  
where~$\sigma$ is a solution of  
\begin{equation}\label{eqn:sig_pert_old}
    -u\sigma = \alpha + \frac{\lambda}{2} \sigma^2.
\end{equation}  
The parameter~$\kappa$ is computed in Ref.~\cite{swartz2023interplay} and is given by  
\begin{equation}
    \kappa = \frac{\int_{-\infty}^\infty (f^{(0)\prime})^2 e^{u^{(0)} z /D_f} \,_2F_1\left(1,\frac{a u^{(0)}}{D_h},1+\frac{a u^{(0)}}{D_h},-e^{z/a}\right) dz}
    {\int_{-\infty}^\infty (f^{(0)\prime})^2 e^{u^{(0)} z /D_f} dz},
\end{equation}  
where~$_2F_1$ is the hypergeometric function.}  
 
\section{Discussion}  
We have explored the effects of spatial anisotropy in a system of two coupled partial differential equations  
previously used to model growth and competition in microbial colonies~\cite{horowitz2019bacterial,lee2022slow, swartz2023interplay,george_chirality_2018}.  
While our analysis employed numerical simulations, perturbative calculations, and pattern selection criteria,  
 simple intuition can also be provided to explain the main results.  
Despite the two-way coupling, we found that the KPZ and FKPP equations contribute to invasion dynamics in largely independent ways.  
This distinction is particularly clear for pulled waves, so we begin by discussing this case.  

For negative~$\alpha$, the invasion velocity~$u$—and thus the position of the sector boundary—is governed solely by the FKPP equation.  
The sector morphology then emerges from the KPZ equation with a source term moving at velocity~$u$,  
which is externally imposed. The solution to this driven KPZ equation is a straight line,  
bounded by the sector boundary on the right and a similar straight line from the left-moving sector boundary on the left.  
Accounting for both edges reproduces the V-shaped dent morphology.  

These dynamics remain largely unchanged as~$\alpha$ becomes positive.  
The invasion speed is still determined by the FKPP equation, while the sector morphology follows  
from a moving source in the KPZ equation. The solution to this driven KPZ equation remains a straight line,  
but now with a negative slope since~$\alpha > 0$.  
This linear segment is again bounded by the sector boundary on the right. However, the left side of the sector now exhibits different behavior.  
Instead of a stationary slope discontinuity at the initial mutant location,  
the central region of the sector transitions into a circular arc~\footnote{Technically a parabolic segment,  
which approximates a circular arc in the limit of small~$\partial_x h$.}.  
This transition is only possible for~$\alpha > 0$ and represents the outward expansion of a faster strain.  
For small~$\alpha$, the horizontal velocity of the circular arc,~$u_{\text{kpz}}$,  
is smaller than the Fisher velocity, $u_0$, so the sector boundary moves ahead of the transition from slope to circular arc.  
This results in the \textit{composite bulge} morphology.  

As~$\alpha$ increases further, the velocity of the circular arc surpasses the Fisher velocity $u_0$.  
At this point, the sector boundary coincides with the transition from a flat to a circular front,  
moving at~$u_{\text{kpz}}$, thus recovering the commonly observed \textit{circular arc} morphology.  
In this regime, invasion is controlled entirely by KPZ dynamics, while the FKPP equation ensures  
that the sector boundary keeps up with the circular arc.  
Indeed, the advection term from the coupling to~$h$ vanishes ahead of the circular arc but  
induces significant advection behind it.  
For nearly isotropic cases where~$\beta \approx \lambda$,  
this advection is strong enough for the mutant strain to keep pace with the circular arc, even when~$s_0 \to 0$.  

The situation changes for~$\beta < \lambda/2$, where the advection velocity, proportional to~$\beta$,  
becomes too weak to maintain pace with the circular arc.  
For sufficiently large~$\alpha$, the combined Fisher and advection velocities become smaller than~$u_{\text{kpz}}$,  
causing the sector boundary to lag behind the transition to a flat front and forming an \textit{escaping bulge} morphology.  
In this regime, the sector boundary and the shock-like singularity move at different velocities:~$u$ and~$c$, respectively.  
Unlike the other three regimes, where only one equation dictates the dynamics,  
both the KPZ and FKPP equations contribute equally to the values of these velocities  
(Eqs.~(\ref{eqn:shock_speed}) and~(\ref{eqn:u_pushed})).  
A key reason for this distinction is that the sector boundary is now entirely within a region of nonzero slope.  

For pushed waves, the same sequence of transitions occurs as~$\alpha$ increases.  
However, unlike pulled waves, the invasion velocity~$u$ now depends on~$\alpha$ for the V-shaped dent and composite bulge morphologies,  
since~$\partial_x h \neq 0$ in the FKPP equation for the bulk of the lateral invasion front.  
Thus, the transitions between morphologies are dictated by the relative values of~$u$ and~$u_{\text{kpz}}$,  
rather than the Fisher velocity $u_0$ and~$u_{\text{kpz}}$.  

\DS{Our analysis relies on several key assumptions.  
First, we assume that colony growth is driven by a thin layer of actively dividing individuals at the leading edge.  
The width of this growth layer is controlled by external factors such as nutrient concentration and diffusivity.  
If individuals within the colony bulk continue to reproduce, our results may no longer apply.  
Additionally, the KPZ framework assumes small front slopes~$\partial_x h$,  
which may not hold for small circular colonies with high curvature or large fitness differences.  
We also assume that interactions mediated by the FKPP equation are sufficiently localized in space.  
If interactions occur over longer length scales—such as via diffusible secreted molecules—  
the FKPP equation may not accurately capture local invasion dynamics.}  

\DS{Several open questions remain for future studies.  
Although we found no striking effects of~$D_h$ on invasion dynamics,  
further exploration of its role in competition and colony morphology is warranted.  
Additionally, negative~$\beta / \lambda$ values could lead to qualitatively different behavior,  
as seen in solid-on-solid growth models and Burgers equations in fluid dynamics.  
Finally, allowing~$\alpha$ and other parameters to depend on~$f$ could introduce new phenomena,  
which may be biologically relevant.  
Despite these limitations, our analysis provides key insights into the interplay between growth and competition  
in expanding populations, particularly within experimental settings relevant to microbial colonies~\cite{lee2022slow}.}  

\DS{In summary, we have characterized how anisotropy influences spatial competition during range expansions.  
While anisotropy can be inferred from non-circular colony shapes, its effect on competition,  
quantified by~$\beta/\lambda$, may not be immediately apparent from growth dynamics alone.  
For pulled waves, weak anisotropy has no effect on population dynamics,  
while for pushed waves, it results only in quantitative changes.  
Thus, detecting anisotropy from sector patterns may be either impossible or require precise measurements.  
Strong anisotropy, however, produces a distinct spatial pattern characterized by  
a shock-like deformation ahead of the sector boundary.  
The emergence of this pattern could serve as a useful signature of anisotropic growth in future studies.}  

\section{Acknowledgements}
D.S. acknowledges support from the MathWorks School of Science Fellowship. H.L. acknowledges support from the Sloan Foundation through grant G-2021-16758. M.K. and D.S. acknowledge support from NSF through grant DMR-2218849.
K.S.K was supported by the NIGMS grant 1R01GM138530-01.

%\end{linenumbers}

\bibliography{Main}{}

@article{van2003front,
  title={Front propagation into unstable states},
  author={Van Saarloos, Wim},
  journal={Physics reports},
  volume={386},
  number={2-6},
  pages={29--222},
  year={2003},
  publisher={Elsevier}
}

@article{van1998three,
  title={Three basic issues concerning interface dynamics in nonequilibrium pattern formation},
  author={van Saarloos, Wim},
  journal={Physics reports},
  volume={301},
  number={1-3},
  pages={9--43},
  year={1998},
  publisher={Elsevier}
}

@article{fisher1937wave,
  title={The wave of advance of advantageous genes},
  author={Fisher, Ronald Aylmer},
  journal={Annals of eugenics},
  volume={7},
  number={4},
  pages={355--369},
  year={1937},
  publisher={Wiley Online Library}
}

@book{murray2002mathematical,
  title={Mathematical biology: I. An introduction},
  author={Murray, James Dickson},
  year={2002},
  publisher={Springer}
}

@misc{kolmogorov1937moscow,
  title={Moscow University Bull},
  author={Kolmogorov, A and Petrovskii, I and Piskunov, N},
  year={1937},
  publisher={Math}
}

@article{hallatschek2007genetic,
  title={Genetic drift at expanding frontiers promotes gene segregation},
  author={Hallatschek, Oskar and Hersen, Pascal and Ramanathan, Sharad and Nelson, David R},
  journal={Proceedings of the National Academy of Sciences},
  volume={104},
  number={50},
  pages={19926--19930},
  year={2007},
  publisher={National Acad Sciences}
}

@article{lee2022slow,
  title={Slow expanders invade by forming dented fronts in microbial colonies},
  author={Lee, Hyunseok and Gore, Jeff and Korolev, Kirill S},
  journal={Proceedings of the National Academy of Sciences},
  volume={119},
  number={1},
  pages={e2108653119},
  year={2022},
  publisher={National Acad Sciences}
}

@article{korolev2012selective,
  title={Selective sweeps in growing microbial colonies},
  author={Korolev, Kirill S and M{\"u}ller, Melanie JI and Karahan, Nilay and Murray, Andrew W and Hallatschek, Oskar and Nelson, David R},
  journal={Physical biology},
  volume={9},
  number={2},
  pages={026008},
  year={2012},
  publisher={IOP Publishing}
}

@article{fife1977approach,
  title={The approach of solutions of nonlinear diffusion equations to travelling front solutions},
  author={Fife, Paul C and McLeod, J Bryce},
  journal={Archive for Rational Mechanics and Analysis},
  volume={65},
  number={4},
  pages={335--361},
  year={1977},
  publisher={Springer}
}

@article{roques2012allee,
  title={Allee effect promotes diversity in traveling waves of colonization},
  author={Roques, Lionel and Garnier, Jimmy and Hamel, Fran{\c{c}}ois and Klein, Etienne K},
  journal={Proceedings of the National Academy of Sciences},
  volume={109},
  number={23},
  pages={8828--8833},
  year={2012},
  publisher={National Acad Sciences}
}

@article{george_chirality_2018,
	title = {Chirality provides a direct fitness advantage and facilitates intermixing in cellular aggregates},
	volume = {14},
	issn = {1553-7358},
	url = {https://journals.plos.org/ploscompbiol/article?id=10.1371/journal.pcbi.1006645},
	doi = {10.1371/journal.pcbi.1006645},
	number = {12},
	urldate = {2023-01-04},
	journal = {PLOS Computational Biology},
	author = {George, Ashish B. and Korolev, Kirill S.},
	month = dec,
	year = {2018},
	note = {Publisher: Public Library of Science},
	pages = {e1006645},
}

@article{horowitz2019bacterial,
  title={Bacterial range expansions on a growing front: Roughness, fixation, and directed percolation},
  author={Horowitz, Jordan M and Kardar, Mehran},
  journal={Physical Review E},
  volume={99},
  number={4},
  pages={042134},
  year={2019},
  publisher={APS}
}

@article{paquette1994structural,
  title={Structural stability and renormalization group for propagating fronts},
  author={Paquette, GC and Chen, Lin-Yuan and Goldenfeld, Nigel and Oono, Y},
  journal={Physical review letters},
  volume={72},
  number={1},
  pages={76},
  year={1994},
  publisher={APS}
}

@article{meerson2011velocity,
  title={Velocity fluctuations of population fronts propagating into metastable states},
  author={Meerson, Baruch and Sasorov, Pavel V and Kaplan, Yitzhak},
  journal={Physical Review E},
  volume={84},
  number={1},
  pages={011147},
  year={2011},
  publisher={APS}
}

@article{mikhailov1983stochastic,
  title={Stochastic motion of the propagating front in bistable media},
  author={Mikhailov, AS and Schimansky-Geier, L and Ebeling, W},
  journal={Physics Letters A},
  volume={96},
  number={9},
  pages={453--456},
  year={1983},
  publisher={Elsevier}
}

@article{rocco2001diffusion,
  title={Diffusion coefficient of propagating fronts with multiplicative noise},
  author={Rocco, Andrea and Casademunt, Jaume and Ebert, Ute and van Saarloos, Wim},
  journal={Physical Review E},
  volume={65},
  number={1},
  pages={012102},
  year={2001},
  publisher={APS}
}

@article{kardar1986dynamic,
  title={Dynamic scaling of growing interfaces},
  author={Kardar, Mehran and Parisi, Giorgio and Zhang, Yi-Cheng},
  journal={Physical Review Letters},
  volume={56},
  number={9},
  pages={889},
  year={1986},
  publisher={APS}
}

@article{kayser2019collective,
  title={Collective motion conceals fitness differences in crowded cellular populations},
  author={Kayser, Jona and Schreck, Carl F and Gralka, Matti and Fusco, Diana and Hallatschek, Oskar},
  journal={Nature ecology \& evolution},
  volume={3},
  number={1},
  pages={125--134},
  year={2019},
  publisher={Nature Publishing Group UK London}
}

@article{giometto2018physical,
  title={Physical interactions reduce the power of natural selection in growing yeast colonies},
  author={Giometto, Andrea and Nelson, David R and Murray, Andrew W},
  journal={Proceedings of the National Academy of Sciences},
  volume={115},
  number={45},
  pages={11448--11453},
  year={2018},
  publisher={National Acad Sciences}
}

@article{hallatschek_life_2010,
	title = {Life at the {Front} of an {Expanding} {Population}},
	volume = {64},
	issn = {1558-5646},
	url = {https://onlinelibrary.wiley.com/doi/abs/10.1111/j.1558-5646.2009.00809.x},
	doi = {10.1111/j.1558-5646.2009.00809.x},
	
	number = {1},
	urldate = {2023-01-09},
	journal = {Evolution},
	author = {Hallatschek, Oskar and Nelson, David R.},
	year = {2010},
	note = {\_eprint: https://onlinelibrary.wiley.com/doi/pdf/10.1111/j.1558-5646.2009.00809.x},
	pages = {193--206},
}

@article{korolev_genetic_2010,
	title = {Genetic demixing and evolution in linear stepping stone models},
	volume = {82},
	url = {https://link.aps.org/doi/10.1103/RevModPhys.82.1691},
	doi = {10.1103/RevModPhys.82.1691},
	number = {2},
	urldate = {2023-01-13},
	journal = {Reviews of Modern Physics},
	author = {Korolev, K. S. and Avlund, Mikkel and Hallatschek, Oskar and Nelson, David R.},
	month = may,
	year = {2010},
	note = {Publisher: American Physical Society},
	pages = {1691--1718},
}

@article{birzu_fluctuations_2018,
	title = {Fluctuations uncover a distinct class of traveling waves},
	volume = {115},
	url = {https://www.pnas.org/doi/abs/10.1073/pnas.1715737115},
	doi = {10.1073/pnas.1715737115},
	number = {16},
	urldate = {2023-01-13},
	journal = {Proceedings of the National Academy of Sciences},
	author = {Birzu, Gabriel and Hallatschek, Oskar and Korolev, Kirill S.},
	month = apr,
	year = {2018},
	note = {Publisher: Proceedings of the National Academy of Sciences},
	pages = {E3645--E3654},
}

@article{giometto_physical_2018,
	title = {Physical interactions reduce the power of natural selection in growing yeast colonies},
	volume = {115},
	url = {https://www.pnas.org/doi/10.1073/pnas.1809587115},
	doi = {10.1073/pnas.1809587115},
	number = {45},
	urldate = {2023-02-09},
	journal = {Proceedings of the National Academy of Sciences},
	author = {Giometto, Andrea and Nelson, David R. and Murray, Andrew W.},
	month = nov,
	year = {2018},
	note = {Publisher: Proceedings of the National Academy of Sciences},
	pages = {11448--11453},
}

@article{swartz2023interplay,
  title={Interplay between morphology and competition in two-dimensional colony expansion},
  author={Swartz, Daniel W and Lee, Hyunseok and Kardar, Mehran and Korolev, Kirill S},
  journal={Physical Review E},
  volume={108},
  number={3},
  pages={L032301},
  year={2023},
  publisher={APS}
}

@article{drossel2000phase,
  title={Phase ordering and roughening on growing films},
  author={Drossel, Barbara and Kardar, Mehran},
  journal={Physical Review Letters},
  volume={85},
  number={3},
  pages={614},
  year={2000},
  publisher={APS}
}

@article{plummer2019fixation,
  title={Fixation probabilities in weakly compressible fluid flows},
  author={Plummer, Abigail and Benzi, Roberto and Nelson, David R and Toschi, Federico},
  journal={Proceedings of the National Academy of Sciences},
  volume={116},
  number={2},
  pages={373--378},
  year={2019},
  publisher={National Acad Sciences}
}

@article{joyce1985molecular,
  title={Molecular beam epitaxy},
  author={Joyce, BA},
  journal={Reports on Progress in Physics},
  volume={48},
  number={12},
  pages={1637},
  year={1985},
  publisher={IOP Publishing}
}

@article{hirschfelder1949theory,
  title={The theory of flame propagation},
  author={Hirschfelder, JO and Curtiss, CF},
  journal={The Journal of Chemical Physics},
  volume={17},
  number={11},
  pages={1076--1081},
  year={1949},
  publisher={American Institute of Physics}
}

@article{dhillon2015critical,
  title={Critical heat flux maxima during boiling crisis on textured surfaces},
  author={Dhillon, Navdeep Singh and Buongiorno, Jacopo and Varanasi, Kripa K},
  journal={Nature communications},
  volume={6},
  number={1},
  pages={8247},
  year={2015},
  publisher={Nature Publishing Group UK London}
}

@article{korolev2015evolution,
  title={Evolution arrests invasions of cooperative populations},
  author={Korolev, Kirill S},
  journal={Physical review letters},
  volume={115},
  number={20},
  pages={208104},
  year={2015},
  publisher={APS}
}

@article{gralka2016allele,
  title={Allele surfing promotes microbial adaptation from standing variation},
  author={Gralka, Matti and Stiewe, Fabian and Farrell, Fred and M{\"o}bius, Wolfram and Waclaw, Bartlomiej and Hallatschek, Oskar},
  journal={Ecology letters},
  volume={19},
  number={8},
  pages={889--898},
  year={2016},
  publisher={Wiley Online Library}
}

@article{beroz2018verticalization,
  title={Verticalization of bacterial biofilms},
  author={Beroz, Farzan and Yan, Jing and Meir, Yigal and Sabass, Benedikt and Stone, Howard A and Bassler, Bonnie L and Wingreen, Ned S},
  journal={Nature physics},
  volume={14},
  number={9},
  pages={954--960},
  year={2018},
  publisher={Nature Publishing Group UK London}
}

@article{gandhi2016range,
  title={Range expansions transition from pulled to pushed waves as growth becomes more cooperative in an experimental microbial population},
  author={Gandhi, Saurabh R and Yurtsev, Eugene Anatoly and Korolev, Kirill S and Gore, Jeff},
  journal={Proceedings of the National Academy of Sciences},
  volume={113},
  number={25},
  pages={6922--6927},
  year={2016},
  publisher={National Acad Sciences}
}

@article{menon2015public,
  title={Public good diffusion limits microbial mutualism},
  author={Menon, Rajita and Korolev, Kirill S},
  journal={Physical review letters},
  volume={114},
  number={16},
  pages={168102},
  year={2015},
  publisher={APS}
}

@article{lavrentovich2014asymmetric,
  title={Asymmetric mutualism in two-and three-dimensional range expansions},
  author={Lavrentovich, Maxim O and Nelson, David R},
  journal={Physical review letters},
  volume={112},
  number={13},
  pages={138102},
  year={2014},
  publisher={APS}
}

@article{nadell2016spatial,
  title={Spatial structure, cooperation and competition in biofilms},
  author={Nadell, Carey D and Drescher, Knut and Foster, Kevin R},
  journal={Nature Reviews Microbiology},
  volume={14},
  number={9},
  pages={589--600},
  year={2016},
  publisher={Nature Publishing Group}
}

@article{korolev2011competition,
  title={Competition and cooperation in one-dimensional stepping-stone models},
  author={Korolev, KS and Nelson, David R},
  journal={Physical Review Letters},
  volume={107},
  number={8},
  pages={088103},
  year={2011},
  publisher={APS}
}

@article{storck2014variable,
  title={Variable cell morphology approach for individual-based modeling of microbial communities},
  author={Storck, Tomas and Picioreanu, Cristian and Virdis, Bernardino and Batstone, Damien J},
  journal={Biophysical journal},
  volume={106},
  number={9},
  pages={2037--2048},
  year={2014},
  publisher={Elsevier}
}

@article{wolf1987wulff,
  title={Wulff construction and anisotropic surface properties of two-dimensional Eden clusters},
  author={Wolf, DE},
  journal={Journal of Physics A: Mathematical and General},
  volume={20},
  number={5},
  pages={1251},
  year={1987},
  publisher={IOP Publishing}
}

@article{devillard1992kinetic,
  title={Kinetic shape of Ising clusters},
  author={Devillard, P and Spohn, H},
  journal={Europhysics Letters},
  volume={17},
  number={2},
  pages={113},
  year={1992},
  publisher={IOP Publishing}
}

@article{hirsch1986anisotropy,
  title={Anisotropy and scaling of Eden clusters in two and three dimensions},
  author={Hirsch, R and Wolf, DE},
  journal={Journal of Physics A: Mathematical and General},
  volume={19},
  number={5},
  pages={L251},
  year={1986},
  publisher={IOP Publishing}
}

@article{li2016gibbs,
  title={Gibbs--Curie--Wulff theorem in organic materials: a case study on the relationship between surface energy and crystal growth},
  author={Li, Rongjin and Zhang, Xiaotao and Dong, Huanli and Li, Qikai and Shuai, Zhigang and Hu, Wenping},
  journal={Advanced Materials},
  volume={28},
  number={8},
  pages={1697--1702},
  year={2016}
}

@book{markov2016crystal,
  title={Crystal growth for beginners: fundamentals of nucleation, crystal growth and epitaxy},
  author={Markov, Ivan Vesselinov},
  year={2016},
  publisher={World scientific}
}

@article{adam2005flowers,
  title={Flowers of Ice—Beauty, Symmetry, and Complexity: A Review of The Snowflake: Winter’s Secret Beauty},
  author={Adam, John A},
  journal={Notices of the AMS},
  volume={52},
  number={4},
  year={2005}
}

@article{langer1989dendrites,
  title={Dendrites, viscous fingers, and the theory of pattern formation},
  author={Langer, JS},
  journal={Science},
  volume={243},
  number={4895},
  pages={1150--1156},
  year={1989},
  publisher={American Association for the Advancement of Science}
}

@article{ben1993snowflake,
  title={From snowflake formation to growth of bacterial colonies: Part I. Diffusive patterning in azoic systems},
  author={Ben-Jacob, Eshel},
  journal={Contemporary Physics},
  volume={34},
  number={5},
  pages={247--273},
  year={1993},
  publisher={Taylor \& Francis}
}

@article{jindal2009theoretical,
  title={Theoretical prediction of GaN nanostructure equilibrium and nonequilibrium shapes},
  author={Jindal, Vibhu and Shahedipour-Sandvik, Fatemeh},
  journal={Journal of Applied Physics},
  volume={106},
  number={8},
  year={2009},
  publisher={AIP Publishing}
}

@article{holmes2015transporting,
  title={Transporting droplets through surface anisotropy},
  author={Holmes, Hal R and B{\"o}hringer, Karl F},
  journal={Microsystems \& Nanoengineering},
  volume={1},
  number={1},
  pages={1--8},
  year={2015},
  publisher={Nature Publishing Group}
}

@article{ahmed2021engineering,
  title={Engineering fiber anisotropy within natural collagen hydrogels},
  author={Ahmed, Adeel and Joshi, Indranil M and Mansouri, Mehran and Ahamed, Nuzhet NN and Hsu, Meng-Chun and Gaborski, Thomas R and Abhyankar, Vinay V},
  journal={American Journal of Physiology-Cell Physiology},
  volume={320},
  number={6},
  pages={C1112--C1124},
  year={2021},
  publisher={American Physiological Society Rockville, MD}
}

@article{stylianopoulos2010diffusion,
  title={Diffusion anisotropy in collagen gels and tumors: the effect of fiber network orientation},
  author={Stylianopoulos, Triantafyllos and Diop-Frimpong, Benjamin and Munn, Lance L and Jain, Rakesh K},
  journal={Biophysical journal},
  volume={99},
  number={10},
  pages={3119--3128},
  year={2010},
  publisher={Elsevier}
}

@article{hong1971magnetic,
  title={Magnetic anisotropy and the orientation of retinal rods in a homogeneous magnetic field},
  author={Hong, Felix T and Mauzerall, David and Mauro, Alexander},
  journal={Proceedings of the National Academy of Sciences},
  volume={68},
  number={6},
  pages={1283--1285},
  year={1971},
  publisher={National Acad Sciences}
}

@article{lam2022asymptotic,
  title={Asymptotic spreading of KPP reactive fronts in heterogeneous shifting environments},
  author={Lam, King-Yeung and Yu, Xiao},
  journal={Journal de Math{\'e}matiques Pures et Appliqu{\'e}es},
  volume={167},
  pages={1--47},
  year={2022},
  publisher={Elsevier}
}

@misc{SI,
  title = {Supplementary Information},
  author = {},
  year = {2025},
  note = {Included as a supplementary document},
}

@misc{aniso_final_repo,
  howpublished = {\url{https://github.com/dancewartz/aniso_final}},
}

@article{mimura2000reaction,
  title={Reaction--diffusion modelling of bacterial colony patterns},
  author={Mimura, Masayasu and Sakaguchi, Hideo and Matsushita, Mitsugu},
  journal={Physica A: Statistical Mechanics and its Applications},
  volume={282},
  number={1-2},
  pages={283--303},
  year={2000},
  publisher={Elsevier}
}

\clearpage

\renewcommand{\thefigure}{S\arabic{figure}}
\renewcommand{\theequation}{S\arabic{equation}}
\onecolumngrid
\textbf{\Large Supplementary Material}
\section{Lattice Implementation}  
\begin{figure}
    \centering
    \includegraphics[width=0.5\textwidth]{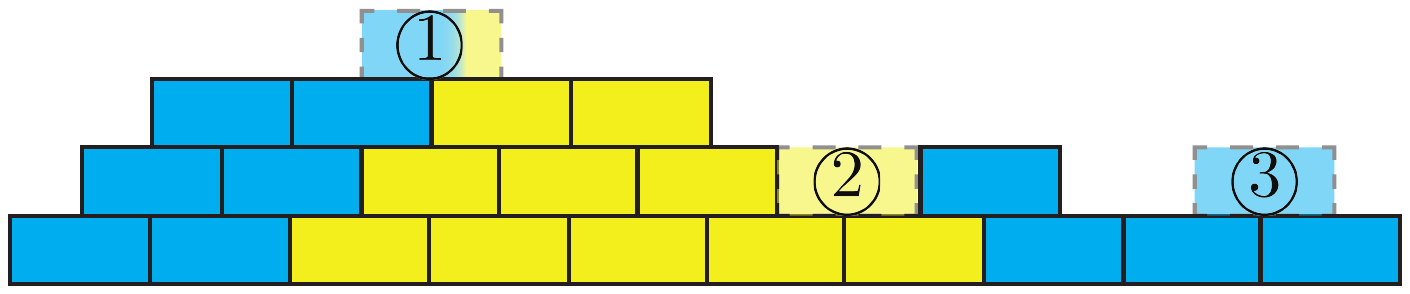}
    \caption{Illustration of the discrete growth model,  
    which prevents overhangs by allowing new growth only when both lower sites are occupied.  
    Even this simple solid-on-solid model produces anisotropic growth.  
    The numbers 1, 2, and~3 correspond to different scenarios determining the color of the newly occupied site.  
    %Further details are provided in the SI.
    }
    \label{fig:cartoon_lattice}
\end{figure}

As a concrete example of anisotropic growth, we consider the solid-on-solid model on a triangular lattice;  
see Ref.~\cite{horowitz2019bacterial} and Fig.~\ref{fig:cartoon_lattice}.  
At each time step, a potential growth site is selected above two already occupied sites,  
ensuring that overhangs do not form.  
The identity of the new cell is then determined by the identities of its two ancestors below,  
as illustrated by sites labeled 1, 2, and 3 in Fig.~\ref{fig:cartoon_lattice}.  

If both ancestors are yellow, a new yellow cell is added with probability~$1 - p_y$.  
If both ancestors are blue, a new blue cell is added with probability~$1 - p_b$.  
If one ancestor is yellow and the other is blue, a new cell is added with probability~$1 - (p_b + p_y)/2$.  
In this mixed-ancestry case, the child cell is blue with probability~$(1 + s/2)$ and yellow otherwise.  

\begin{figure*}
    \centering
    \includegraphics[width=\textwidth]{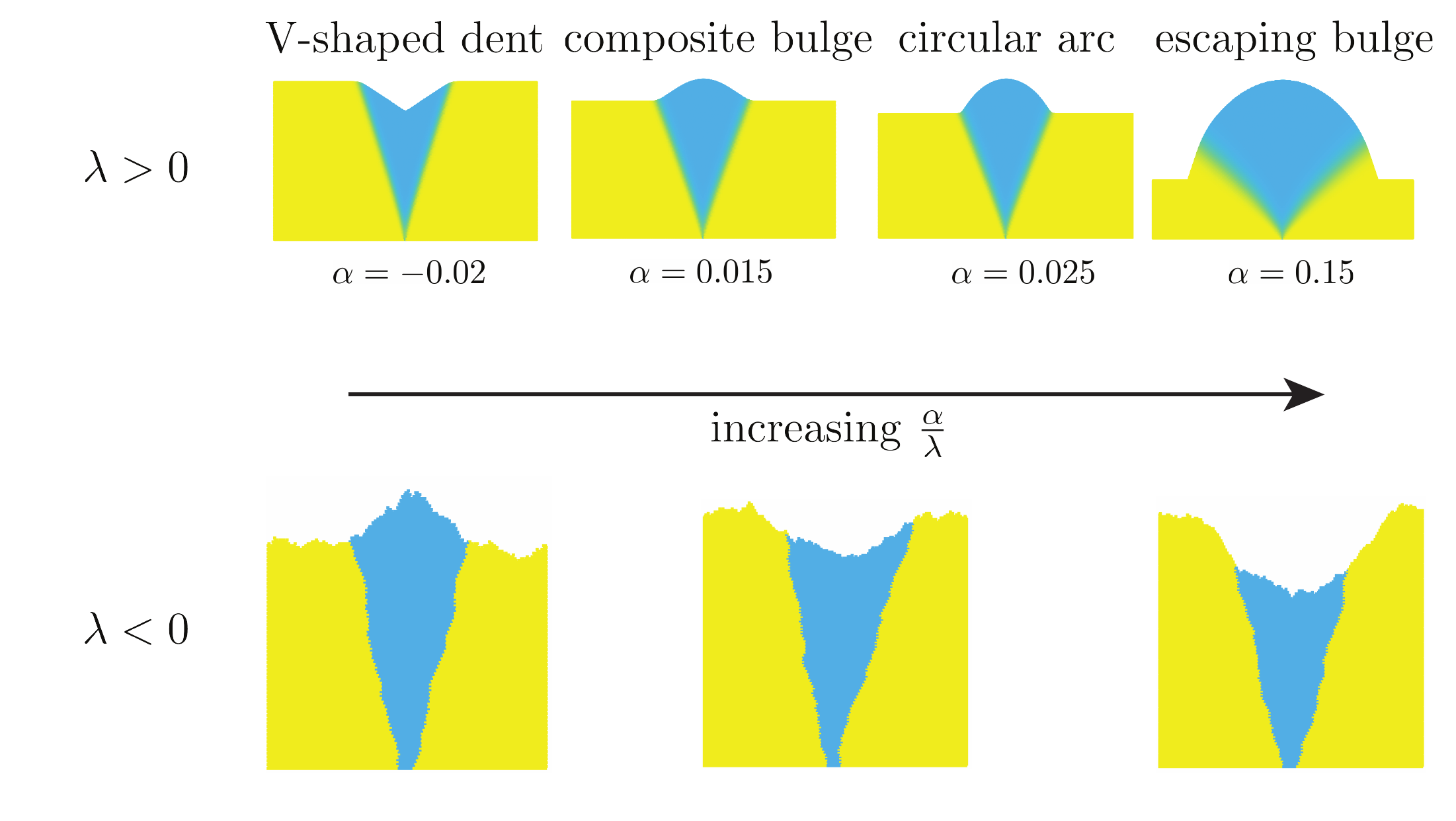}
    \caption{Comparison of predicted morphologies (top panel, numerical solutions of Eqns (1) and (2)) with results from stochastic simulations (bottom panel).  
    The bottom panel demonstrates that similar results hold in stochastic numerical simulations;  
    see Fig.~\ref{fig:cartoon_lattice}.  
    Previous work has shown that this solid-on-solid lattice model exhibits a negative~$\lambda$,  
    so to compare with the top panel, one must apply the transformation  
    $(h, \beta, \alpha) \to (-h, -\beta, -\alpha)$, see \cite{horowitz2019bacterial} for details.  
    With this transformation, the leftmost image corresponds to the V-shaped dent,  
    while the rightmost image represents the escaping bulge  
    (where the sector boundary lies within a tilted region).  
    The middle image corresponds to either the composite bulge or the circular arc.  
    Distinguishing between these two cases requires simulations at much larger system sizes.  
    \textbf{Simulation Parameters:}  
    - \textit{Top row}:~$D_f = 1$, $D_h = 1$, $v_0 = 0.08$, $\lambda = 20$, $s_0 = 0.25$, and~$\beta = 5$.  
    - \textit{Bottom row}: Lattice model parameters~$(p_b, p_y, s)$ from left to right:  
      $(0, 0.3, 0.5)$, $(0.17, 0.3, 0.5)$, and~$(0.3, 0, 0.5)$.}
    \label{fig:lattice_morphologies}
\end{figure*}

The parameters~$p_y$ and~$p_b$ control the effective growth rates of the strains  
and can be related to~$v_0$ and~$\alpha$ in Eq.~(2).  
The additional parameter~$s$ describes the effects of local competition and is related to~$s_0$ in Eq.~(1).  
Typical spatial patterns produced by this model are shown in Fig.~\ref{fig:lattice_morphologies}.  

\section{Other variants of the two-dimensional mechanistic model}

In the main text, we introduced a pair of mechanistic two-dimensional equations  
that reproduce all previously studied morphologies (dent, composite, and circular)  
and demonstrated that removing spatial isotropy leads to the emergence of the escaping bulge morphology,  
as predicted by the coupled FKPP-KPZ equations.  

A key aspect of this model is the distinction between rescalable and non-rescalable anisotropy.  
In our model, non-rescalable anisotropy is introduced by allowing the diffusion coefficients  
to vary across spatial directions and between species.  
Our choice to use a linear diffusion operator was arbitrary,  
and different biological scenarios may require alternative formulations.  
To assess the generality of the escaping bulge morphology,  
we performed simulations on variants of the two-dimensional model discussed in the text,  
incorporating density-dependent diffusion:  
\begin{equation}
    \frac{\partial n_i(x,y,t)}{\partial t} = 
    r_i(1-n)(n-K) \left(1 + \sum_j a_{ij} \frac{n_j}{n} \right) n_i \nonumber 
    + \frac{\partial}{\partial x} \left(D_{ix}(n) \frac{\partial n_i}{\partial x}\right) 
    + \frac{\partial}{\partial y} \left(D_{iy}(n) \frac{\partial n_i}{\partial y}\right).
\end{equation}
Here, diffusivity still varies in magnitude between strains and directions,  
but its density dependence remains identical across both strains and spatial directions.  
We tested two commonly used forms of density-dependent diffusion,  
which model cooperative interactions such as mechanical pressure buildup in the colony:  
$D(n) \propto n$ and~$D(n) \propto n(1-n)$.  

\begin{figure}
    \centering
    \includegraphics[width=\linewidth]{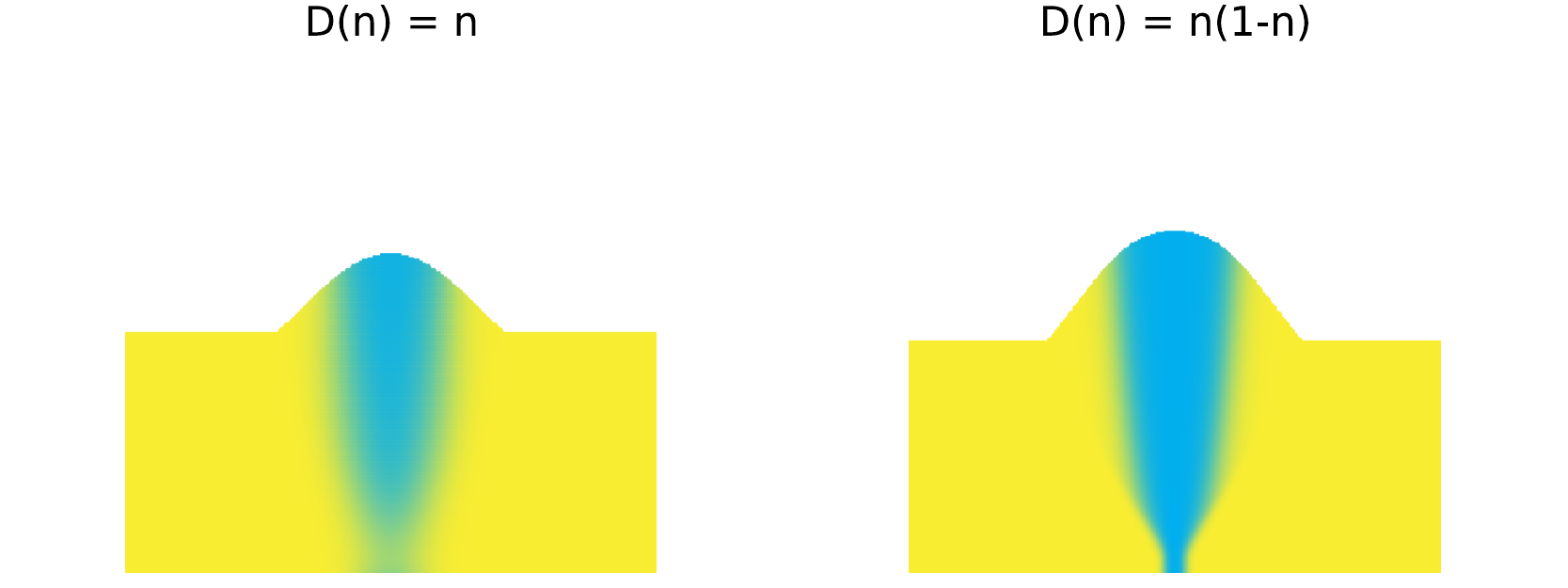}
    \caption{Escaping bulge morphology appears during anisotropic growth for different density-dependent diffusion coefficients. For the left, parameters are $r_1=r_2=0.35, a_{11} = 0.6, a_{21}=0.6, a_{12} = a_{22} = 0, D_{1x}=D_{1y}=D_{2y}=0.03, D_{2x}=0.05$. For the right panel, parameters are $r_1=r_2=0.35, a_{11} = 0.8, a_{21}=0.8, a_{12} = a_{22} = 0, D_{1x}=D_{1y}=D_{2y}=0.03, D_{2x}=0.05$.} 
    \label{fig:compare_density_dependence}
\end{figure}

Example morphologies are shown in Fig.~\ref{fig:compare_density_dependence}.  

\begin{figure}
    \centering
    \includegraphics[width=\textwidth]{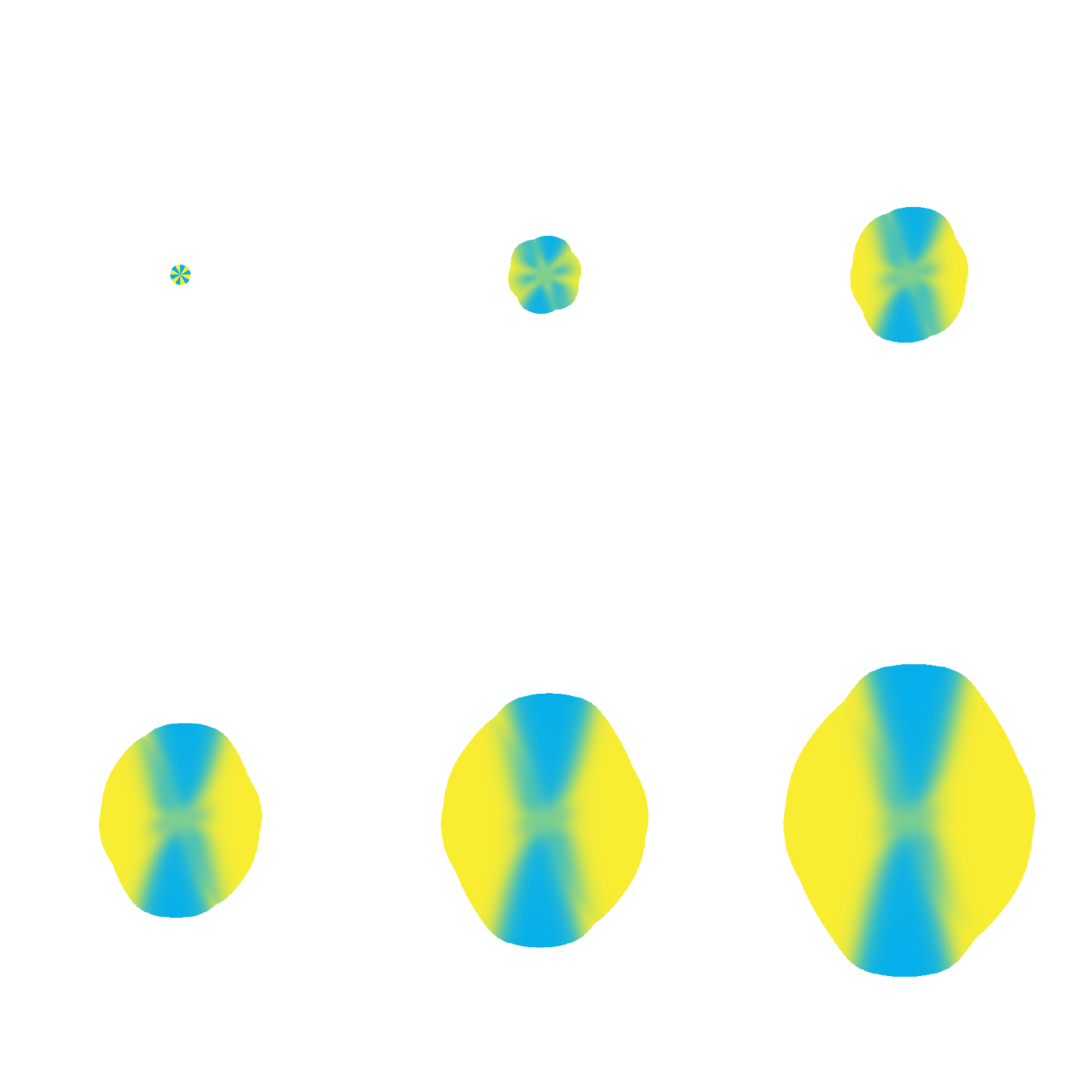}
    \caption{Under anisotropic growth, sector boundaries drift to align with the axis of slowest growth. Here, the yellow strain has a larger diffusion coefficient in the horizontal direction, while the blue has a growth rate advantage. The initial condition shown on the left is a small circular colony with radial sectors. We see that as time progresses from left to right the sectors that are oriented horizontally favor the yellow strain while the vertical sectors favor the blue, eventually producing an escaping bulge morphology oriented vertically. Simulations were produced using the two-dimensional reaction diffusion model discussed in the main text. Parameters are $r_1=r_2=1.5, a_{11}=2.4, a_{21}=2.38, a_{12}=a_{22}=0, D_{1x}=D_{1y}, D_{2y}=0.005, D_{2x}=0.007$}
    \label{fig:non_vertical_anisotropic_sectors}
\end{figure}

\begin{figure}
    \centering
    \includegraphics[width=\linewidth]{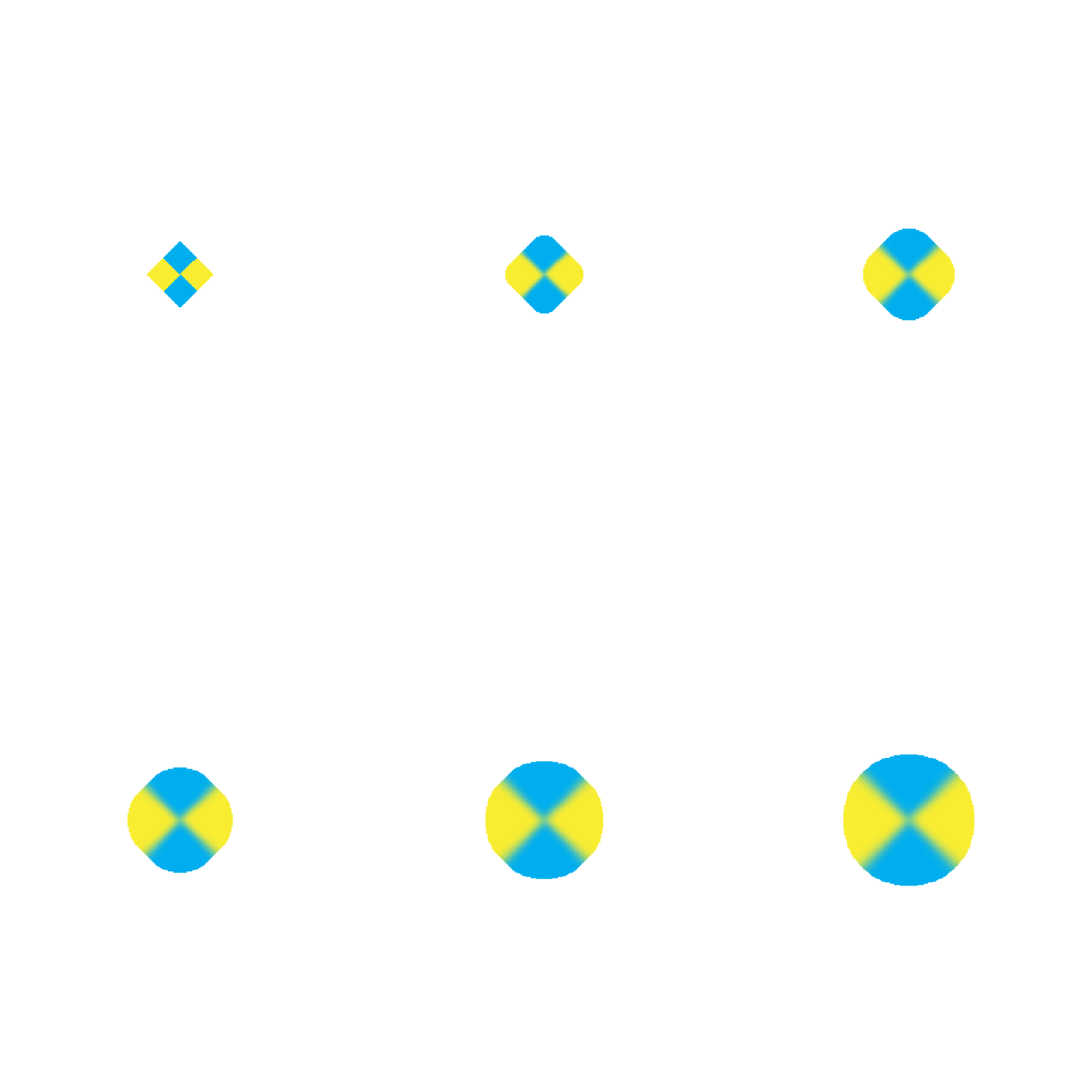}
    \caption{Under neutral isotropic growth sector boundaries remain perpendicular to the expansion front at all times $\epsilon=0$. Shown are multiple snapshots of a colony of two neutral strains with a diamond initial condition for the colony shape. We see that the sector boundaries remain straight with time and always make a 90 degree angle with the expansion front. Simulations were produced using the two-dimensional reaction diffusion model discussed in the main text. Parameters are $r_1=r_2=1.5, a_{11}=a_{21}=a_{12}=a_{22}=0, D_{1x}=D_{1y}=D_{2y}=D_{2x}=0.005$.}
    \label{fig:enter-label}
\end{figure}

\begin{figure}
    \centering
    \includegraphics[width=\linewidth]{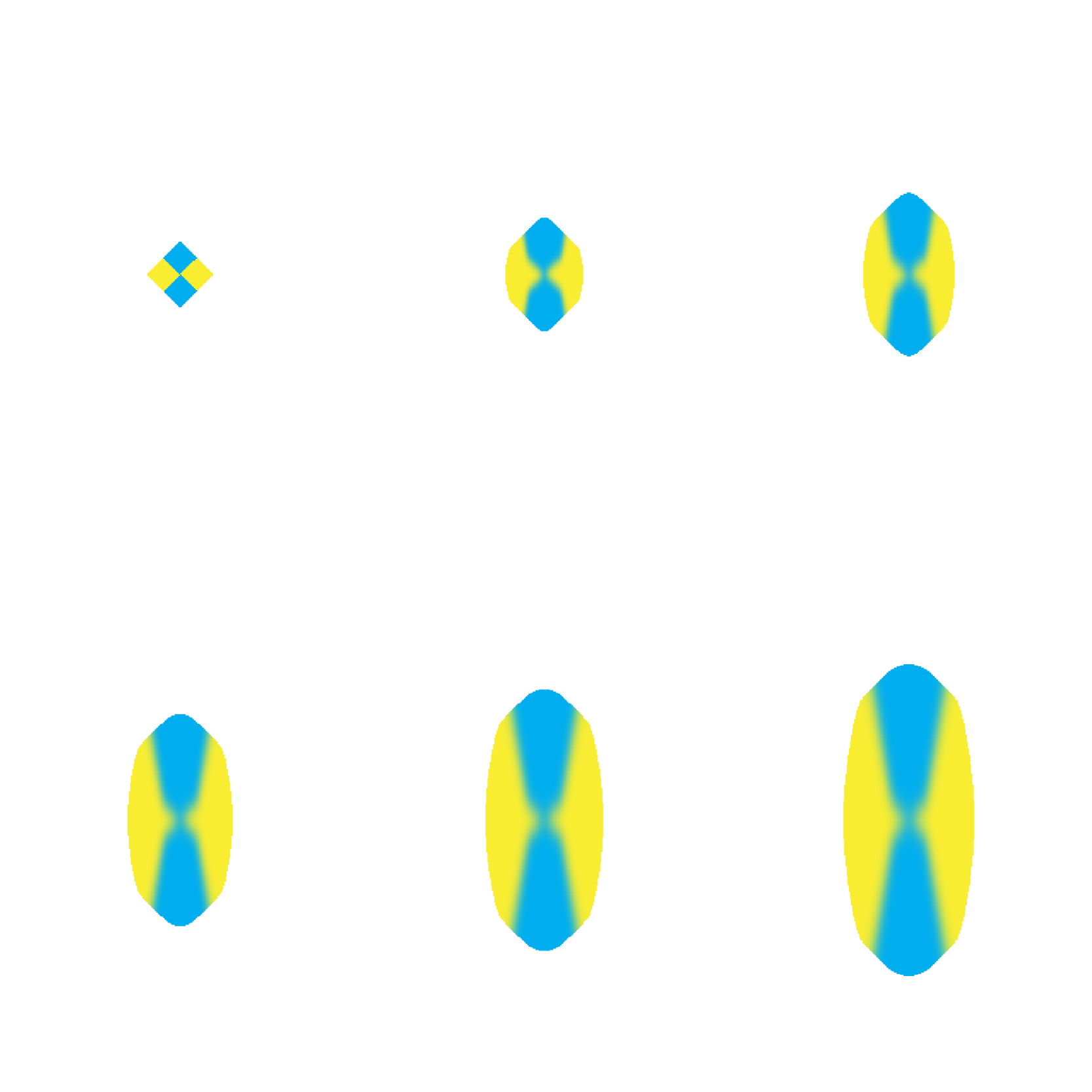}
    \caption{Under neutral anisotropic growth sector boundaries are not necessarily perpendicular to the expansion front at all times, $\epsilon \neq 0$. Shown are multiple snapshots of a colony of two neutral strains with a diamond initial condition for the colony shape. Simulations were produced using the two-dimensional reaction diffusion model discussed in the main text. Parameters are $r_1=r_2=1.5, a_{11}=a_{21}=a_{12}=a_{22}=0, D_{1x}=D_{2x}= 0.005, D_{1y}=D_{2y}=0.03$.}
    \label{fig:enter-label}
\end{figure}

\begin{figure*}
    \centering
    \includegraphics[width=\textwidth]{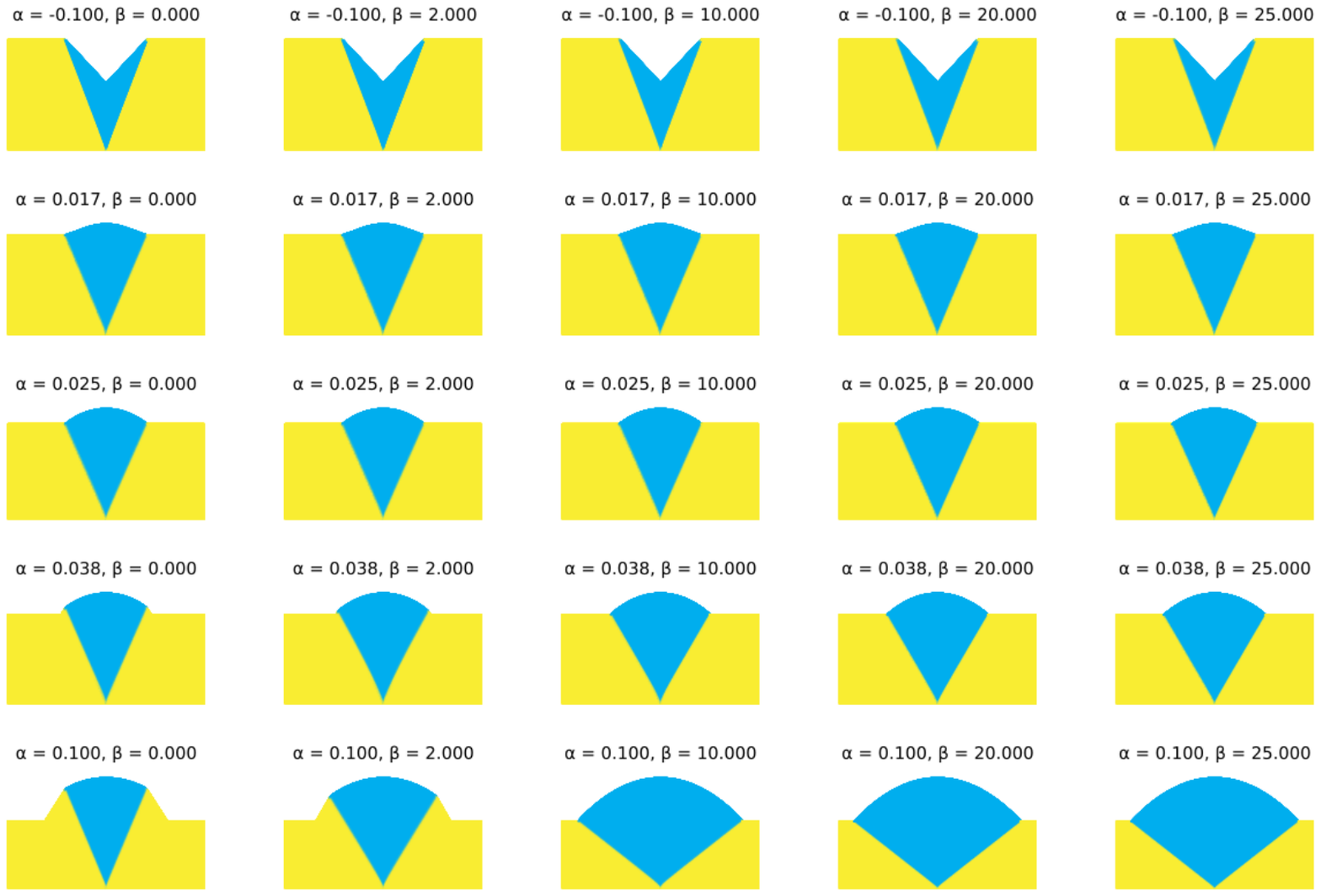}
    \caption{Collection of sample morphologies as a function of both the expansion speed difference $\alpha$ and advection coupling $\beta$ \textcolor{blue}{generated by numerically solving Eqns (1) and (2)}. We see that the escaping bulge morphology only appears for large $\alpha$ and small $\beta$. Parameters are $D_f=D_h=1, v_0=0.15, \lambda=20, s_0=0.25$ and the Fisher dynamics are that of a pulled wave.}
    \label{fig:enter-label}
\end{figure*}

\end{document}